\documentclass[manuscript,screen,nonacm]{acmart}

\usepackage{etoolbox}

\newcommand{\smallish}{\fontsize{8.5pt}{9.5pt}\selectfont}
\AtBeginEnvironment{quote}{\smallish\itshape}
\newcommand{\iquote}[1]{{\smallish``\textit{#1}''}}

\usepackage{booktabs}
\usepackage{tabularx}
\usepackage{array}

\AtBeginDocument{%
  }

\acmISBN{978-1-4503-XXXX-X/2018/06}

\usepackage{xcolor}
\definecolor{amber}{RGB}{255,191,0}
\definecolor{joypink}{RGB}{199,21,133}

\newcommand{\model}{\textit{Adaptive Gratitude Practice Model}}

\newcommand{\Situation}{(\Situation)}
\newcommand{\pattern}{(\pattern)}
\newcommand{\Pattern}{(\Pattern)}
\newcommand{\practice}{(\practice)}
\newcommand{\Practice}{(\Practice)}
\begin{document}

\title[Beyond Counting Blessings: Tracing the Evolution of Gratitude Practices and Technology Needs]{Beyond Counting Blessings: Tracing the Evolution of Gratitude Practices and Technology Needs}




\author{Joy Qiuyue Zhong}
\orcid{0009-0006-4328-2329}
\affiliation{%
  \department{Manning College of Information \& Computer Sciences}
  \institution{University of Massachusetts Amherst}
  \city{Amherst}
  \state{Massachusetts}
  \country{United States}
}
\email{qzhong@umass.edu}

\author{Jeongah Lee}
\orcid{0000-0002-5714-5521}
\affiliation{%
  \department{Manning College of Information \& Computer Sciences}
  \institution{University of Massachusetts Amherst}
  \city{Amherst}
  \state{Massachusetts}
  \country{United States}
}
\email{jeongahlee@umass.edu}

\author{Drishti Goel}
\orcid{0009-0000-6713-9240}
\affiliation{%
  \department{Siebel School of Computing and Data Science}
  \institution{University of Illinois Urbana Champaign}
  \city{Champaign}
  \state{Illinois}
  \country{United States}
}
\email{drishti4@illinois.edu}

\author{Violeta J. Rodríguez}
\orcid{0000-0001-8543-2061}
\affiliation{%
  \department{Department of Psychology}
  \institution{University of Illinois Urbana-Champaign}
  \city{Champaign}
  \state{Illinois}
  \country{United States}
}
\email{vjrodrig@illinois.edu}

\author{Dong Whi Yoo}
\orcid{0000-0003-2738-1096}
\affiliation{%
  \department{Luddy School of Informatics, Computing, and Engineering}
  \institution{Indiana University Indianapolis}
  \city{Indianapolis}
  \state{Indiana}
  \country{United States}
}
\email{dy22@iu.edu}

\author{Koustuv Saha}
\orcid{0000-0002-8872-2934}
\affiliation{%
  \department{Siebel School of Computing and Data Science}
  \institution{University of Illinois Urbana-Champaign}
  \city{Urbana}
  \state{Illinois}
  \country{United States}
}
\email{ksaha2@illinois.edu}

\author{Ravi Karkar}
\orcid{0000-0003-1467-4439}
\affiliation{%
  \department{Manning College of Information \& Computer Sciences}
  \institution{University of Massachusetts Amherst}
  \city{Amherst}
  \state{Massachusetts}
  \country{United States}
}
\email{rkarkar@cs.umass.edu}

\renewcommand{\shortauthors}{Zhong et al.}



\begin{abstract}

Gratitude technologies support well-being by prompting reflection on what people appreciate. But gratitude does not serve the same purpose in every circumstance: as life situations change, so does what people seek from it, and whether it feels appropriate at all. To understand how technology can adapt to and support such shifts, we conducted retrospective, artifact-elicitation interviews with 17 adults who had practiced gratitude for one to fifteen years. Participants' appraisals of their situations shaped what they needed, yielding six recurring practice patterns, including a boundary where gratitude felt forced. We contribute the \model{}, which explains how appraisals shifted even within the same life situation, how participants adapted activities, modalities, and rhythms, lapsed under competing demands or emotional unreadiness, and resumed when gratitude again felt useful. Additionally, we derive design implications for situated support, self-understanding through past records, and relational care with changing life situations.

\end{abstract}




\maketitle
\section{Introduction}

In the wake of global crises and rising social isolation~\cite{abdalla2026loneliness, santomauro2026updated, zhang2025global}, health organizations encourage gratitude practice as an everyday way to support well-being and social connection. The U.S. Centers for Disease Control and Prevention (CDC) encourages people to keep gratitude journals, thank others, and offer help as ways to manage stress~\cite{cdc2026managingstress, cdcGratitudeWorks}. These are ways of practicing gratitude: recognizing benefits received from others and appreciating what is valuable in life, including during difficulty~\cite{emmons2003counting, wood2010gratitude, algoe2012find}. The U.S. Surgeon General's resources similarly include expressing gratitude among activities intended to strengthen relationships~\cite{haglandconnection}. 

In recent years, AI-driven well-being practice has increasingly moved onto personal devices\cite{song2025exploreself, kim2024diarymate, kim2024mindfuldiary}. Smartphone and web systems now scaffold gratitude technology through daily prompts, notification-triggered reminders, context-sensitive cues, and shared social feeds \cite{ghandeharioun2016kind, isaacs2013echoes, zhang2024people, bunn2022gogratitude, heckendorf2019efficacy, lei2025exploring}. Yet prior work highlights persistent challenges. Effects of gratitude interventions are modest and heterogeneous, and depend heavily on dosage, timing, and fit with the individual \cite{davis2016thankful, cregg2021gratitude, huston2025understanding, choi2025meta}. Repeated reflection can also become less effective through habituation \cite{lyubomirsky2005pursuing, sheldon2006increase}. Adherence is a further barrier. Qualitative accounts of gratitude journaling echo this, with people reporting uncertainty about what "counts," repetition of the same entries, and the practice drifting from a reflective act into an obligation to be discharged \cite{bhattacharjee2024actually, ko2021nursing, huston2025understanding}. These experiences reflect broader challenges in sustaining engagement with mental-health apps and reflective tools, particularly when the effort of recording exceeds its perceived value~\cite{baumel2019objective, epstein2016beyond}.
Together, these challenges raise a question for HCI: how can technology support gratitude practice in ways that fit people's everyday lives?

Prior HCI research has examined several aspects of this fit, including environmental cues for prompting, routines and preferences for tailoring interaction, and emotional and social circumstances shaping people's experiences~\cite{ghandeharioun2016kind, bhattacharjee2024actually, lee2024cultivating, tang2022co, kaltenhauser2026connected}. This fit may also change over time as life situations reshape what people seek from gratitude and whether they feel ready to practice. Consider an illustrative scenario: during a stressful working week, someone uses an evening gratitude prompt to notice small and valued moments, such as colleague's help or an enjoyable conversation. This reflection may help them cope with daily stress by reminding them of the support and positive experiences in their life. After losing their job or a loved one, however, the same prompt might feel pressuring when they need space to express distress before reflecting on what they appreciate or pause their practice. Prior research supports the importance of such questions of fit: the benefits of positive activities depend partly on their fit with the person~\cite{lyubomirsky2013simple}, and gratitude exercises can feel difficult or invalidating during recent or ongoing hardship~\cite{huston2025understanding}.

Broader well-being research motivates attention to these life situations, it examines associations between life-event attributes and well-being and calls for frameworks that account for individual differences and situational appraisals~\cite{saha2026life,saha2025mental}. This perspective suggests that understanding the role of gratitude technology in people's lives requires considering both their circumstances and how they interpret them. 

However, how people adapt gratitude practice as these situations change remains underexplored. We therefore examine gratitude practice across life situations, encompassing significant events as well as ongoing or recurring circumstances, and sought to investigate ``\textbf{How do changing life situations shape how people practice gratitude, and how can we design technology that dynamically accommodates these situational shifts?}''. Understanding how people adapt practice within these situations can inform technology that accommodates changing needs over time. In particular, we need to understand how they practice gratitude in everyday life, how it evolves and is maintained, and how technology supports or constrains it. Such an understanding can guide decisions about when and how technologies should offer support as people's needs change. Accordingly, our work is guided by the following RQs:

\begin{itemize}
\item \textbf{RQ1:} How does gratitude practice adapt to people’s changing life situations and needs over time?

\item \textbf{RQ2:} What role does and can technology play in supporting gratitude practice across life situations and over time?
\end{itemize}



To develop this understanding, we conducted a retrospective qualitative study with 17 adults who reported practicing gratitude for one to fifteen years. During artifact-elicitation interviews, participants revisited selected gratitude records to discuss their practices and the circumstances surrounding them. These records supported recall and contextual explanation, while interview transcripts formed the main analytic corpus. We analyzed the interview transcripts  using reflexive thematic analysis \cite{braun2019reflecting, braun2022conceptual}.

Our analysis suggests that participants’ appraisals of their situations shaped what they sought from gratitude and whether it felt appropriate. They adapted practice patterns as appraisals and needs changed, and adjusted activities, modalities, or rhythms to fit practical conditions and preferences. They reported lapses during competing demands or emotional unreadiness and resumed when gratitude again felt useful or manageable. Perceived value and broader life commitments provided reasons to continue or resume. We bring these findings together in the \model{} (Overview in figure~\ref{fig:study_overview}). The model illustrates how needs vary across individuals in similar life situations and within individuals over time. It provides a framework for technology to support people’s changing needs and the ways they adapt gratitude practice over time. 

This work makes three contributions:

\begin{itemize}

\item \textbf{Theoretical Contribution}: \model{} explains how people's life situation appraisals shape what they need from gratitude practices, which in turn inform practice patterns, while practical conditions and preferences shape their implementation. 
how experienced values and broader life commitments inform maintenance and resumption, and how limited capacity and emotional unreadiness leads people to lapse, with practice remaining available for resumption.

\item \textbf{Empirical Evidence}: We identified \textit{six} recurring practice patterns describing what people seek from gratitude and when it fits their situated needs, including : Making Space for Difficult Emotions, making sense of hardship, appreciating what remained present and available, connecting effort and progress with a valued future, keeping meaningful experiences available over time, and recognizing and Extending Care.

\item\textbf{Design Recommendations}: We identify opportunities for technology to provide situated and evolving gratitude practice, including tailoring support to people's life situation appraisals, respecting emotional unreadiness, supporting self-understanding through past records, and facilitating the seeking and expression of care.

\end{itemize}

\section{Background and Related Work}
We situate the study in research on gratitude as a well-being practice (Section~\ref{sec:gratitude_wellbeing}), technology support for gratitude in context (Section~\ref{sec:gratitude_tech}), and long-term engagement in personal informatics (Section~\ref{sec:gratitude_pi}).

\subsection{Gratitude as an Everyday Well-being Practice: Benefits and Boundaries}
\label{sec:gratitude_wellbeing}

Rising rates of stress, anxiety, and loneliness following global disruptions such as the COVID-19 pandemic have contributed to a mental health burden now affecting people worldwide~\cite{abdalla2026loneliness, santomauro2026updated, zhang2025global}. Governments and health organizations have increasingly prompted gratitude as everyday practices people can adopt on their own~\cite{cdc2026managingstress, cdcGratitudeWorks,haglandconnection}. Gratitude is broadly defined as recognizing benefits received from others and appreciating what is valuable in life, including in the face of hardship ~\cite{wood2010gratitude}.

HCI and psychology researchers have studied gratitude mainly through structured exercises such as journaling, counting blessings, and gratitude letters, generally finding modest and variable improvements to well-being
~\cite{emmons2003counting,davis2016thankful, choi2025meta}, while expressing appreciation directly to others has also been shown to strengthen relationships~\cite{algoe2012find,kumar2018undervaluing}. 

These benefits depend partly on the conditions of practice. Positive activity research identifies timing, variety, dosage, motivation, and person--activity fit as influences on outcomes~\cite{lyubomirsky2013simple}, while cross-cultural studies examine how cultural context shapes responses~\cite{layous2013culture}. Gratitude may also feel inappropriate. Interviews with mental health professionals and health psychology researchers suggest that adversity can make appreciation meaningful, while recent or ongoing hardship can make gratitude exercises difficult or invalidating~\cite{huston2025understanding}. Interpersonal expectations also matter: two vignette studies found that stronger expectations of repayment increased indebtedness and reduced gratitude~\cite{watkins2006debt}. These studies highlight the importance of understanding both whether gratitude fits a person's circumstances and how it is practiced.

Prior research work identifies benefits and conditions that shape the suitability of gratitude practice, and much of this evidence concerns prescribed gratitude exercises evaluated over a limited period. They offer less insight into how people incorporate gratitude into their lives over time, including when they find it useful, what they seek from it, and how they adapt their practices as their circumstances change. We address these questions through an empirical study of how people navigate these decisions across their long-term gratitude practice histories, informing technology support for changing needs and situations.

\subsection{Context-Sensitive Technology Support for Well-being and Gratitude}
\label{sec:gratitude_tech}

Research on well-being technologies examines how support can respond to people's states, goals, and everyday contexts. Just-in-time adaptive interventions provide a framework for tailoring support to changing states and contexts~\cite{nahum2016just,suh2024toward}, with receptivity-aware approaches using contextual information to adapt intervention timing~\cite{mishra2021detecting,mishra2024exploring}. Research on conversational support also examines the needs arising in specific settings, including workplace emotional labor, caregiving, and mental health self-management~\cite{das2025ai,shi2026mapping,saha2026ai,yoo2026ai}. \textit{CASEbot} provides an example of incorporating personal context into system interaction by eliciting users' goals and everyday constraints to personalize health self-experiments~\cite{ishita2026casebot}. These studies illustrate approaches to considering both when people can engage with support and how it can accommodate their circumstances.

Within gratitude research, technologies examines how support is delivered, how people interact with it, and how emotional and social contexts shape its use. The \textit{Kind and Grateful} app uses location changes, social proximity, and physical activity to trigger prompts~\cite{ghandeharioun2016kind}. Other studies investigate speech assistants~\cite{helgert2022you}, augmented reality~\cite{bunn2022gogratitude}, and chatbots~\cite{lee2024cultivating} to support gratitude practice. Related positive psychology work explores tangible interfaces for the Three Good Things intervention~\cite{siriaraya2024happy}. User-centered studies identify opportunities for personalization, reminders, multimedia entries, reflective prompts, mood labeling, and low-effort participation~\cite{blabst2018new,bhattacharjee2024actually}, while comparisons of self-directed and guided journaling examine how guidance relates to engagement and outcomes~\cite{zhao2026self}. This work offers design options for accommodating different preferences and ways of practicing.

Emotional and social contexts further shape how gratitude support is experienced. Chatbot research highlights the coexistence of gratitude and negative emotions~\cite{lee2024cultivating}, while systems for coworkers, care partners, and communities embed appreciation in ongoing relationships~\cite{tang2022co,hsu2023gratibot,zhang2024people,makri2020can,walsh2023optimal,lei2025exploring,kaltenhauser2026connected,shi2025balancing}. Reflective archives also support revisiting meaningful experiences, although recording alone does not ensure reflection~\cite{carvalho2019emojar,baumer2015reflective,cho2022reflection}. Cross-cultural and practitioner-facing research further examines variation in gratitude intervention outcomes and suitability~\cite{choi2025meta,huston2025understanding}. These studies draw attention to the feelings, relationships, and past experiences that give gratitude practice meaning.

Together, these studies explain how delivery, interaction, and emotional and social contexts shape well-being and gratitude practice. what remains largely unaddressed is how people's changing life situations shape what they seek from gratitude and how they adapt their practices accordingly. We examine these relationships over time to inform technology that responds to changes in gratitude's relevance and purpose within everyday life.

\subsection{Gratitude as an Understudied Practice in Personal Informatics}
\label{sec:gratitude_pi}

Personal informatics research offers useful concepts for understanding long-term engagement. Lived informatics situates tracking within everyday life and includes lapsing and resuming as part of engagement~\cite{rooksby2014personal,epstein2015lived}. People also adapt tracking methods and reconsider goals as their needs and circumstances change, including during life transitions~\cite{ayobi2018flexible,feron2022transitions,ekhtiar2025changing}. Research on technology-mediated reflection further examines how personal concerns, available information, and timing shape opportunities for reflection~\cite{bentvelzen2021technology}.

Continued engagement can also have costs. Reflection may contribute to rumination or distress~\cite{eikey2021beyond,luo2025reflecting}, while discontinuation can reflect changing priorities, movement between tools, or a sense that a tool has served its purpose~\cite{epstein2016beyond,baumer2018departing}. These perspectives motivate examining what changes in engagement mean within people's lives.

Gratitude recording offers a context for extending these perspectives because it involves documenting valued experiences that people may later revisit and reflect on. This raises questions about how such records remain meaningful as people's lives change, and what adaptation, lapse, and resumption mean within the broader practice of gratitude. We draw on personal informatics research to examine these questions through people's long-term practice histories and explore their needs for technological support.

\section{Method}
We conducted a retrospective qualitative study with 17 adults who had practiced gratitude for one to fifteen years. We first describe our recruitment and study design (Section~\ref{sec:design}), reveal our interview procedure (Section~\ref{sec:interview}), and finally our reflexive thematic analysis approach (Section~\ref{sec:rta}).

\subsection{Study Design and Recruitment}
\label{sec:design}

For this study, gratitude practice referred to intentional, recurring efforts to notice, reflect on, record, or communicate appreciation. ~\cite{wood2010gratitude,emmons2003counting, kumar2018undervaluing} Eligible participants were at least 18 years old, had practiced gratitude for at least six months, and could share and discuss at least ten gratitude entries spanning six months or more. Their broader practices often also included other forms of journaling and reflection. Requiring records across time supported our aim of exploring changes in practice.

We recruited participants through university channels, social media, online gratitude communities, and word-of-mouth sampling. From eligible applicants, we purposively selected 17 participants to vary in age, life circumstances, and practice history, with practice lengths ranging from one to 15 years. Table~\ref{tab:participants} summarizes participant demographics and practice duration.

The study was approved by the Institutional Review Board at our institution. All participants gave informed consent and received \$30. We replaced identifying details in transcripts, quotations, and artifacts with participant IDs. Participants chose which records to share and could skip questions or withdraw an artifact from discussion.

Before the interviews, we conducted a formative review of public iOS App Store reviews of 18 gratitude-related applications. 
This review provided an initial view of how people used these applications and raised questions that short, cross-sectional reviews could not answer, particularly about changes within the same person over time. These questions informed our interview protocol. The reviews were not part of the analytic corpus reported in this paper.

\begin{table}[t]
\centering
\caption{Participant demographics, gratitude practice duration, and records shared in interviews.} 
\label{tab:participants}

\begingroup
\footnotesize
\setlength{\tabcolsep}{3pt}

\begin{tabularx}{\linewidth}{
    @{}
    >{\raggedright\arraybackslash}p{0.03\linewidth}
    >{\raggedright\arraybackslash}p{0.06\linewidth}
    >{\raggedright\arraybackslash}p{0.03\linewidth}
    >{\raggedright\arraybackslash}p{0.08\linewidth}
    >{\raggedright\arraybackslash}p{0.16\linewidth}
    >{\raggedright\arraybackslash}p{0.18\linewidth}
    >{\raggedright\arraybackslash}p{0.055\linewidth}
    >{\raggedright\arraybackslash}X
    @{}
}
\toprule
\textbf{ID} &
\textbf{Gender} &
\textbf{Age} &
\textbf{Race/ Ethnicity} &
\textbf{Education} &
\textbf{Occupation} &
\textbf{Practice Length} &
\textbf{Sources of Records Brought to Interviews} \\
\midrule

P01 & Female & 29 & African  & Bachelor's &
Virtual assistant & 2 yrs &
Paper/phone journal, Gratitude App journal  \\

P02 & Male & 25 & African  & Bachelor's &
Electrical engineer & 10 yrs &
Paper/phone journal, Gratitude App journal \\

P03 & Female & 25 & African  & Bachelor's &
Data analyst & 2 yrs &
Paper/phone journal \\

P04 & Female & 26 & Asian & Master's &
PhD student & 1 yr &
Paper journal \\

P05 & Female & 41 & Asian & Master's &
Therapist & 10 yrs &
Paper/phone journal\\

P06 & Female & 25 & Asian & Bachelor's &
Master's student & 3 yrs &
Paper/phone journal \\

P07 & Female & 30 & Multiracial & Bachelor's &
UI/UX designer & 1.5 yrs &
Paper/phone journal, Gratitude App journal \\

P08 & Female & 26 & Asian & Bachelor's &
PhD student & 1 yr&
Paper journal, Gratitude App journal \\

P09 & Female & 22 & Asian & Bachelor's &
PhD student & 5 yrs &
Paper journal  \\

P10 & Female & 22 & White & Bachelor's &
College student & 5 yrs &
Phone journal \\

P11 & Male & 52 & White & Master's &
Contract specialist & 1.5 yrs&
Paper journal, Gratitude app journal \\

P12 & Female & 55 & White & Bachelor's &
CPA & 8 yrs&
Gratitude app journal \\

P13 & Male & 68 & White & Doctorate &
Semi-retired professor & 10 yrs&
Gratitude app journal \\

P14 & Male & 47 & White & Doctorate &
Consultant, board member& 15 yrs&
Paper journal,Gratitude app journal \\

P15 & Female & 65 & White & Bachelor's &
Realtor & 8 yrs&
Gratitude app journal\\

P16 & Female & 54 & White & Bachelor's &
Education specialist & 5 yrs&
Gratitude app journal \\

P17 & Male & 35 & Asian & Bachelor's &
Software developer & 9 yrs &
Gratitude app journal \\

\bottomrule
\end{tabularx}
\endgroup
\end{table}

\subsection{Interview Procedure}
\label{sec:interview}
We conducted individual semi-structured interviews over Zoom, each lasting approximately one hour. Participants brought gratitude records, including paper journals with drawings and doodling, app entries, phone notes and walked us through selected entries from different periods to explain the circumstances surrounding them. We followed up on changes and important moments and periodically summarized participants' accounts to invite clarification or correction.

Interviews were recorded, transcribed, checked against recordings, and de-identified. 
Artifacts primarily supported recall and provided context, while transcripts formed the main analytic corpus. The Selected anonymized gratitude entries participants gave us permission to reproduce appear in \hyperref[app:entry]{Appendix}.

\subsection{Reflexive Thematic Analysis}
\label{sec:rta}
We analyzed the interviews using Braun and Clarke's reflexive thematic analysis (RTA) ~\cite{braun2006using,braun2019reflecting,braun2021one,braun2022conceptual}. We examined how participants made sense of their experiences, recognizing that their accounts reflected their life circumstances and the interview context.

The first author, who led the interviews and analysis, brought an HCI background and prior participation in online gratitude communities. This familiarity supported rapport and understanding of community practices but may also have directed attention toward gratitude's benefits. Interactions before informed consent were excluded from the study data. The first author reread transcripts, revisited recordings, and wrote analytic memos. Initial coding focused on why participants practiced gratitude, how they practiced, their experiences with tools, what support it sustain, how do they feel before and after gratitude practice, and periods of continuation and lapse. The first and second authors independently coded four transcripts and discussed their interpretations. The first author then coded the remaining transcripts, while the second reviewed them alongside developing codes to identify alternative interpretations and overlooked details. These discussions informed revisions to codes and candidate themes; we did not calculate coding agreement.

We compared accounts across participants and across different periods within each participant's history. These comparisons drew attention to how participants understood their situations and what they sought from gratitude. As the analysis developed, literature on appraisal, gratitude, coping, emotion regulation, and personal informatics helped us interpret these relationships. These concepts informed interpretation without determining the initial codes. We used diagrams to examine how life situation appraisals, situated psychological needs, practice patterns, implementation, and engagement related over time.

The broader team's expertise in psychology, personal informatics, and well-being informed discussions of the themes and diagrams. We examined our assumptions by revisiting supporting and contrasting accounts, particularly when gratitude felt unavailable, burdensome, or emotionally inappropriate. These accounts helped clarify the limits of gratitude's relevance and the meanings of adaptation, lapse, and resumption. Team discussions, evidence selection, and writing further refined the themes, model, construct definitions, and scope of our claims.

\section{Findings}

\begin{figure}[t]
  \centering
  \includegraphics[width=\linewidth]{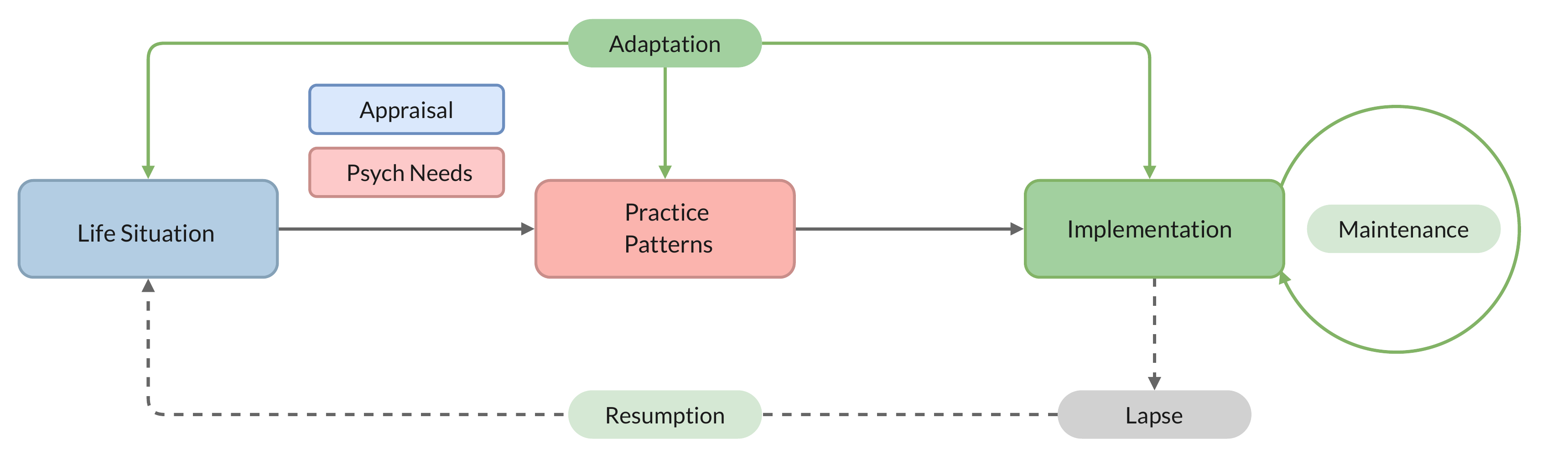}
  \caption{Overview of the \model{} illustrating how gratitude practice evolves with changing life situations and situated psychological needs. People's appraisals shape what they seek from gratitude, informing the practice patterns they draw on and how they implement them. When the implementation fits the need and circumstance, it is repeated as maintenance; when the fit breaks down, people adapt it or lapse, and practice remains available to be resumed in a different form.}
  \label{fig:study_overview}
\end{figure}

Through reflexive thematic analysis, we developed the \model{} (model overview - figure~\ref{fig:study_overview}) to describe how gratitude practice evolves with people's changing life situations and needs. We further elaborate the core model and its alternate pathways (detailed figure~\ref{fig:study_detail}). The core model shows how participants' appraisals of their life situations shaped their situated psychological needs, which informed their practice patterns. Participants implemented these patterns through activities, modalities, and rhythms that fit their practical considerations and preferences. When established practice patterns and implementation continued to fit current needs and capacity, participants maintained their practice. Over time, however, changes in life situations or how participants appraised them could reshape their psychological needs and, in turn, their practice patterns, while changes in practical conditions and preferences could affect implementation. The alternate pathways therefore describe how participants responded to these shifts through adaptation, lapse, and resumption.

To address RQ1, we examine when and why people engage in gratitude practice (Section~\ref{sec:relevance}), how people engaged in gratitude practice  (Section~\ref{sec:implementation}), and how their practices evolved over time (Section~\ref{sec:continuity}). We address RQ2 by investigating participants' views on how technology could support their long-term practices (Section~\ref{sec:tech-support}).

\subsection{Situated Psychological Needs Shaped Whether and How Gratitude Became Relevant}
\label{sec:relevance}

Whether gratitude felt appropriate and how participants engaged with it depended on what they needed psychologically in a particular life situation. These needs were shaped by how participants understood their circumstances. We call these participant-reported meanings and judgments \textit{life situation appraisals}, drawing on appraisal theories that relate people's responses to a situation's significance for their well-being, goals, and coping~\cite{lazarus1984stress,moors2013appraisal}. These appraisals shaped the support participants sought, which we term \textit{situated psychological needs}. Similar situations could raise different needs, and different situations could raise similar needs. Appraisals and needs could also change while a life situation remained ongoing.

We identified six recurring \textit{practice patterns} through which participants responded to situated psychological needs (Table~\ref{tab:gratitude-patterns}). \textit{Distress processing} marked a boundary of gratitude's relevance: participants sometimes needed to acknowledge and express overwhelming or unresolved emotions before gratitude felt appropriate. Five gratitude-related patterns described other forms of support: \textit{reinterpretation} - making hardship more understandable; \textit{recalibration} - broadening attention toward what remained present and available; \textit{reorientation} - turning toward a valued future; \textit{preservation} - keeping meaningful experiences available over time; and \textit{relational care} - appreciating care received and expressing and showing care for others.

These patterns could occur independently, coexist within a reflection, or shift as appraisals changed. Participants sometimes combined distress and gratitude, expressed distress before later engaging gratitude, or engaged gratitude directly. Reinterpretation, recalibration, and reorientation were especially evident in responses to difficulty, while preservation and relational care emphasized valued experiences and relationships. These overlapping patterns describe recurring responses to situated needs without implying a fixed sequence.

\begin{table}[t]
\centering
\caption{Six recurring patterns of situated gratitude practice}
\label{tab:gratitude-patterns}

\begingroup
\scriptsize
\setlength{\tabcolsep}{4pt}
\setlength{\parindent}{0pt}
\setlength{\parskip}{0pt}
\renewcommand{\arraystretch}{1.10}

\newcommand{\ghead}[1]{%
    {\footnotesize\bfseries #1\par}%
    \vspace{3pt}%
}
\newcommand{\gcase}[1]{#1\par}
\newcommand{\gquote}[2]{%
    \noindent\textup{#1:}\enspace\textit{``#2''}\par
}

\begin{tabularx}{\linewidth}{
    @{}
    >{\hsize=0.70\hsize\linewidth=\hsize
      \raggedright\arraybackslash}X
    >{\hsize=0.95\hsize\linewidth=\hsize
      \raggedright\arraybackslash}X
    >{\hsize=1.15\hsize\linewidth=\hsize
      \raggedright\arraybackslash}X
    >{\hsize=1.15\hsize\linewidth=\hsize
      \raggedright\arraybackslash}X
    >{\hsize=1.05\hsize\linewidth=\hsize
      \raggedright\arraybackslash}X
    @{}
}
\toprule
{\footnotesize\bfseries Pattern} &
{\footnotesize\bfseries Life situations} &
{\footnotesize\bfseries Appraisal} &
{\footnotesize\bfseries Situated psychological needs} &
{\footnotesize\bfseries Illustrative quote} \\
\midrule
\addlinespace[3pt]

\ghead{Distress processing}
&
\ghead{Loss and strain}
\gcase{Bereavement; caregiving crises}
\gcase{Illness; breakups; homesickness}
&
\ghead{Distress demands attention}
Pain feels overwhelming, and gratitude may feel inappropriate or insufficient.
&
\ghead{Space for distress}
Difficult emotions need acknowledgment and expression without judgment.
&
\gquote{P09}{I started journaling whenever I used to feel very overwhelmed. I can write any hateful comment in my journal, and nobody would judge me. So it was definitely a stress reliever.}
\\
\addlinespace[4pt]
\midrule
\addlinespace[4pt]

\ghead{Reinterpretation}
&
\ghead{Personal setbacks}
\gcase{Breakups; friendship loss}
\gcase{Unemployment; work difficulties}
&
\ghead{Meaning or self-worth is challenged}
Hardship raises questions about what happened and what it says about the self.
&
\ghead{Making sense of hardship}
A more constructive understanding can acknowledge difficulty alongside learning or resilience.
&
\gquote{P15}{What can I take from that situation where that guy slammed his door in my face and left me shaking? how can I grow from it?}
\\
\addlinespace[4pt]
\midrule
\addlinespace[4pt]

\ghead{Recalibration}
&
\ghead{Adversity and daily pressures}
\gcase{Illness; bereavement}
\gcase{Financial scams; demanding work}
&
\ghead{Difficulty dominates attention}
What is wrong or missing overshadows what remains available.
&
\ghead{Balance and grounding}
Recognizing what remains meaningful or sufficient restores a sense of balance.
&
\gquote{P09}{I have so much now that I have to be proud of, rather than holding on to what happened}
\\
\addlinespace[4pt]
\midrule
\addlinespace[4pt]

\ghead{Reorientation}
&
\ghead{Change and competing demands}
\gcase{Bereavement; career transitions}
\gcase{Addiction recovery}
\gcase{Work and family demands}
&
\ghead{Direction or progress feels uncertain}
A valued future feels unclear or difficult to move toward.
&
\ghead{Agency and competence}
Meaningful choices and progress offer direction and confidence to move forward.
&
\gquote{P03}{I'm using gratitude as a way to set direction. It's forward-looking, focused on growth, self-respect, long-term goals}
\\
\addlinespace[4pt]
\midrule
\addlinespace[4pt]

\ghead{Preservation}
&
\ghead{Meaningful moments and transitions}
\gcase{Everyday joys; milestones}
\gcase{Children growing up}
\gcase{Loss of loved ones or pets}
&
\ghead{Valued memories may fade}
Meaningful experiences may become harder to recall as life changes.
&
\ghead{Continuity with what matters}
Memories need to remain accessible for revisiting or passing on.
&
\gquote{P16}{I was afraid I was gonna forget, like, lose all my memories of him}
\\
\addlinespace[4pt]
\midrule
\addlinespace[4pt]

\ghead{Relational care}
&
\ghead{Encounters with care}
\gcase{Support during difficulties}
\gcase{Community acceptance}
\gcase{Others' kindness or vulnerability}
&
\ghead{Care affirms connection}
Kindness signals being valued, while vulnerability reveals opportunities to offer care.
&
\ghead{Assurance and contribution}
Connection involves feeling supported and finding ways to thank, encourage, or help others.
&
\gquote{P16}{they just opened their hearts to me, my experience was accepted as truthful and valid and worthy.}
\vspace{3pt}
\gquote{P15}{it's giving me the opportunity to show my compassion and support for them.}
\\
\addlinespace[3pt]
\bottomrule
\end{tabularx}

\vspace{5pt}
\begin{minipage}{\linewidth}
\scriptsize
\textit{Note.}
Situations are illustrative; appraisals and needs are analytic summaries.
Similar situations can evoke different appraisals, and patterns can
coexist or shift over time. Distress processing includes moments
when gratitude feels inappropriate or unavailable.
\end{minipage}

\endgroup
\end{table}


\subsubsection{Pattern 1 - Distress Processing: Making Space for Difficult Emotions}

\label{sec:boundary}

When participants appraised a situation as overwhelming, unresolved, or not yet understandable, their immediate need was to acknowledge what they felt and give those emotions an outlet. At these moments, turning toward positive experiences could feel forced and premature. Participants needed space to process anger, fear, grief, or confusion without pressure to resolve the situation. Gratitude could remain unavailable when it did not address this need. P12 described this boundary during a period when her son experienced repeated overdoses and her husband was seriously ill:
\begin{quote}
``My son overdosed a few times, and I had to perform CPR on him. \ldots I finally reached out and got some help, and the counselor suggested a gratitude practice. I tried a written journal and that didn't work.\ldots
I was really on my own, and no matter the amount of gratitude journaling or exercise I did, it was not gonna make that better.''---P12
\end{quote}

Gratitude could not change the crisis she was facing. Still, writing helped her release some of what P12 had been carrying: \iquote{Writing it down makes me let go of it. \ldots I don't need to think about this anymore.} P16 described a similar role during grief without pressure to move on: \iquote{You just need to get things out, so they're not weighing on your heart.} These accounts illustrate a mismatch between the situated psychological need and what gratitude could offer at that moment: the unresolved crisis first demanded acknowledgment. Writing responded to this need by giving distress a bounded place outside the self, without requiring participants to change how they felt about it.

For some participants, expressing distress later created room for a different view of the situation. P11 explained that after getting difficult thoughts out, his mindset could change:\iquote{Once I get it out, \ldots [I] start looking at there's gotta be something good in there.} P09 initially vented anger, but later found that venting alone left her angry and was \iquote{not directing [the emotions] in a way that's helpful.}

These cases mark an important boundary for gratitude: participants sometimes needed to process distress before they felt ready to reappraise what they had experienced. Some then explored what the hardship had taught them, revealed about them, or enabled them to value. We describe this as \textit{reinterpretation}.

\subsubsection{Pattern 2 - Reinterpretation: Making Sense of Hardship}

Hardship prompted meaning-making when participants struggled to reconcile an experience with their existing beliefs and sense of self. These situations raised a need for coherence: understanding why an event affected them and what it meant for their lives. This need often arose during unemployment, relationship ruptures, and existential uncertainty, as experiences changed how participants saw themselves. Participants sought learning and alternative interpretations without necessarily viewing the event itself as positive. P13 explained: \iquote{gratitude practice is for cognitive reframing. I call it more like the positive re-interpretation.} P11 use simple word to explain the meaning of reinterpretation:\iquote{everything works for you not against you.} 

For P03, a breakup and prolonged unemployment raised doubts about worth and abilities. Her gratitude reflections helped her separate the hardship from a broad negative judgment about herself (Appendix: Figure~\ref{fig:P03}):
\begin{quote}
\iquote{I literally asked myself, what is it teaching me? What strength is this building? \ldots I realized I was actually being resilient and just trying consistently. That shift changed how I saw myself\ldots I couldn’t be grateful for not having a job, but I could be grateful for my resilience\ldots I don't turn bad into good, I just look for growth inside.}—P03 
\end{quote}

P03's account shows the boundary of reinterpretation. She did not redescribe unemployment as a good event. She found resilience and growth she valued in how she had responded to it. P09 similarly used gratitude to recognize that she had grown from earlier hardship instead of continuing to define herself through what had happened.

Reinterpretation changed the meaning assigned to a difficult experience or to the self within it. It did not require finding a benefit in every hardship, particularly during ongoing stress or irreparable loss. In such situations, participants sometimes needed to prevent difficulty from becoming the only visible part of life. They broadened their attention toward what remained present while the meaning of the hardship remained unchanged. We call this \textit{recalibration}.

\subsubsection{Pattern 3 - Recalibration: Appreciating What Remained Present and Available.}Recalibration became relevant when difficulty or uncertainty narrowed participants' attention to what was wrong, missing, or likely to go wrong. During illness, grief, anticipated demands, or feelings of scarcity, meaningful and supportive parts of life became less visible. Many participants described this recognition, as P11 explained:\iquote{Even when I was angry, there was always something to be happy about if you look for it.} P12 described being \iquote{mired in the moment} and \iquote{circling all of the negative thoughts}. Participants needed to notice what remained present and available. We call this \textit{recalibration}: restoring broader, more balanced attention while the meaning of the difficulty could remain unchanged. P14 described this imbalance at the start of a demanding workday:
\begin{quote}
\iquote{I would sort of start my day with, what are the 3 things that can go wrong today, and what are the 27 things that I need to do today, a terrible way of starting your day, because then you focus on the negative, or the potentially negative.}---P14
\end{quote}

Recalibration also helped when disappointment threatened to dominate a moment. After losing a large amount of money shortly before Christmas, P07 appreciated being with her family, having good health, and sharing Christmas dinner. Noticing these alongside the loss kept it from defining the whole occasion. P10 similarly grounded attention in \iquote{what is in front of you} when facing an uncertain future.

\subsubsection{Pattern 4 - Reorientation: Connecting Effort and Progress with a Valued Future}

Participants used gratitude to connect present choices, effort, and progress with a valued future. Loss, divorce, unemployment, unwanted habits, uncertainty, caregiving stress, and competing demands could make that future difficult to imagine or pursue. Appreciating their available choices and accomplishments helped participants clarify their direction and recognize their capacity to move toward it. We call this \textit{reorientation}: identifying what to pursue and recognizing everyday actions as meaningful steps toward it. After her husband's death, P16 described choosing to continue living while carrying her grief:

\begin{quote}
\iquote{I have a lot of living that I still want to do \ldots I'm choosing to carry that grief forward with me. \ldots The life that we had together, and the life that I still have to live, and I don't have to give up one for the other.}---P16
\end{quote}

Appreciating everyday progress also connected participants' actions to the lives they wanted. During recovery from alcohol-related difficulties, P11 celebrated leaving a store without buying alcohol as \iquote{a little win.} He also tracked workouts and appreciated his persistence (Appendix: Figure~\ref{fig:P11}). Amid work demands, P12 reviewed her journal to recognize accomplishments and \iquote{get back to the goals,} including volunteering, making art, donating blood, and learning French. P14 connected gratitude with daily intentions to be kind, patient, calm, and loving toward his children.

\subsubsection{Pattern 5 - Preservation: Keeping Meaningful Experiences Available over Time}

The possibility of forgetting valued experiences made keeping a record meaningful to participants. Everyday joys, important milestones, and moments of connection were worth preserving because their details could fade and life could change. This appraisal raised a need to retain a memory with people and periods of life that mattered. \textit{preservation} addressed this need by keeping meaningful experiences available for later appreciation, often through records. P06 explained:

\begin{quote}
\iquote{There are so many positive things happening in our life, but we just forget about it. If we keep a track in the journal, or we document these things, those things really made us happy when we look back. \ldots We need to do it immediately, because after 2 or 3 days, we just forget that.}---P06
\end{quote}

P10 similarly recorded \iquote{small energy moments,} such as unexpected conversations and shared meals( \ref{fig:P10}), alongside \iquote{big energy moments,} such as graduation. These records could also serve others. P13 downloaded ten years of gratitude posts to leave to his children as a \iquote{legacy.} Preservation thus sustained personal connections with the past and made that history available to others.

\subsubsection{Pattern 6 - Relational Care : Recognizing and Extending Care}

Moments of kindness and acceptance reassured participants that they mattered to others and sometimes prompted a wish to extend care themselves. This reassurance was especially meaningful when they felt alone, ashamed, or lost. Remembering support or having difficult experiences accepted met a need to feel connected and cared for, which we call \textit{relational assurance}. Appreciating kindness received or witnessed could also make participants want to thank, encourage, or help others, which we call \textit{prosocial connection}.

P02 recorded acts of kindness to remember both the experience and the person, explaining that these memories reminded him that \iquote{people really do care about me.}(\ref{fig:P02}) Feeling lost after leaving a job he loved to travel, P17 wrote a long thank-you message to a friend. Recalling their happy shared experiences moved him from \iquote{feeling really miserable to feeling really amazing.}

Appreciating care also prompted participants to consider how they could support others. P07 explained:

\begin{quote}
\iquote{I started to reflect on all of the good deeds that I have received and those little things that I can do for other people to make them happy. We share food, visit [people] and help. I was grateful for the fact that I can do [those things]}---P07
\end{quote}

P16 used reactions and thoughtful comments to acknowledge community members and validate their experiences because \iquote{it's important that people feel seen.} P12 shared her struggles \iquote{to encourage people around me.}

Together, these cases show how gratitude met relational needs when participants felt alone, ashamed, or lost. It helped them remember that they mattered to others and gave them ways to thank, encourage, or support someone else. For some, remembering the care they had received became a reason to pass that care forward.

\subsection{Considerations and Preferences Shaping Implementation: Adapting Activity, Modality, and Rhythm}
\label{sec:implementation}

Participants carried out the practice patterns described in Section~\ref{sec:relevance} in ways shaped by their practical considerations and preferences. We describe these choices as \textit{practice implementation}, comprising activity, modality, and rhythm. \textit{Activity} refers to what participants did, \textit{modality} to the medium or interface they used, and \textit{rhythm} to when and how often they practiced. 
How participants put these patterns into practice depended on their everyday routines and preferences. P14 illustrated this flexibility:
\begin{quote}
\iquote{In the morning, I was usually sort of writing down bullets, that were related to overarching things \ldots In the evening, when I did the gratefulness practice, sort of wrapping up my day, then I was usually writing being grateful for something that happened during the day, then it became quite repetitive \ldots I switched from Sort of writing it down in my physical little booklet to going to the app.}—P14
\end{quote}

\subsubsection{Activity: Choosing How to Reflect and Who to Practice With} \textbf{Participants shaped activities around their preferences for guidance or open expression and the opportunities their social settings offered for practicing with others}. Activity refers to what participants did in their practices. Activities varied along two related dimensions: how they were structured and who took part. Participants moved between guided and open-ended activities, and among personal, interpersonal, and collective practice.

\paragraph{Choosing Between Guided and Open-Ended Reflection} Participants practiced through different types of structure and focus such as short-lists, answered reflective prompts, longer journal entries, letters to themselves, and positive self-talk writing or free expressing writing. P14 commonly listed three to five bullet points to shift attention to positivity,  while prompts gave P12 a starting point when open-ended journaling had not worked. Open-ended activities gave participants more control over what to include. P06's evening journal combined events, feelings,stressful thoughts, and positive insights:
\begin{quote}
\iquote{At the end of the day, what things happened with me? What were the significant events? How I am feeling, what [are] my positive insights? \ldots I also keep track of my negative thoughts, or what made me stressed.}—P06
\end{quote}

This format allowed P06 to acknowledge distress and gratitude within the same activity. P09 similarly used free writing for an early entry she called a \iquote{distress rant}, then later wrote a letter to herself to encourage herself:\iquote{six years ago, you wouldn’t have believed that you were here.} Many participants mixed expressions of distress with gratitude-related entries, and some later revisited earlier distress and found something useful in it. 

Sometimes, participants used inward practice such as breathing, meditation, sensory attention, and prayer to notice what was present or connect their experiences to spiritual beliefs. P10 practiced gratitude through mindful observation:\iquote{loved staring at the rain and watching the drop slide down the cracked open window.} P13's spiritual practice directed attention toward what he called the “big basics,” including “life, breath, food, warm place to sleep" as well as \iquote{taking the time to stop and watch a garden do what gardens do.}

Activity structure therefore shaped what participants could include, while the absence of a written entry did not by itself show whether the broader practice had continued.

\paragraph{Organizing Shared Practice Within Relationships and Groups} Activities also varied in their social reach. A personal practice could remain private, become a message to another person, or develop into a coordinated group activity. P17 sent a thank-you message to a friend recalling experiences they had shared. In online communities, P14 wrote detailed comments on others' gratitude posts, while P16 responded to vulnerable disclosures by sharing experiences of suffering and appreciation.

Collective activities required participants to coordinate what people attended to, expressed, and shared.For example, P16 incorporated paired gratitude reflection and group sharing into staff meetings:
\begin{quote}
\iquote{[We] talk with a partner about someone you are grateful for working with this week \ldots or a goal that you're grateful for accomplishing this week. Then we'll share out for the good of the group.}---P16
\end{quote}

P05 and P14 similarly organized gratitude writing and discussion with their families. In P13's recovery community, members invited newcomers to recognize one day of sobriety as an achievement and celebrated it together. These activities made appreciation something others could receive, affirm, and continue.

Activity was therefore not a fixed gratitude exercise: participants changed its structure and focus, kept it inward or expressed it outwardly, and moved between personal and shared practice as their purposes and situations changed.

\subsubsection{Modality: Matching Expressive Preferences and Practical Constraints}

\textbf{Participants selected and combined modalities to support their preferred ways of expressing experiences while accommodating tool availability, time, and energy.} We use \textit{modality} broadly to describe how an activity was carried out and expressed, including paper-based and digital forms and written, visual, and spoken expression. Participants selected, shifted between, and combined modalities as their purpose, availability, time, and energy changed.

\paragraph{Matching Modalities to the Purpose of Expression.}

Modalities shaped the pace, structure, and expressive possibilities of an activity. Participants therefore selected them according to how they wanted to engage with and express an experience. P03 contrasted writing by hand with using a phone:
\begin{quote}
\iquote{When I journal in a paper notebook, my entries tend to be more emotional and more detailed, because writing by [hand] slows me down, so I sit with my feelings longer. \ldots When I use my phone, my entries are usually shorter [and] more immediate.It feels more convenient, but less immersive.}---P03
\end{quote}

For P03, handwriting slowed the activity and supported sustained attention, while the phone enabled shorter and more immediate expression. Digital modalities could also support extended engagement through different forms of structure. App prompts gave P12 a starting point when open-ended journaling had not worked, while P07 used guided apps to remain present and reflect. P17 used a Google Document for longer entries when working through difficult thoughts.

Participants also used different modalities to expand what they could express. P11 combined writing with drawings, doodles, and stickers, describing paper as an outlet to \iquote{be creative and do whatever I want to do. It just frees me.}, while P15 used voice input because hearing the words made the expression feel more powerful.

These cases show that paper and digital modalities did not serve fixed patterns. Their fit depended on the pacing, structure, space, and expressive options provided by a particular implementation.

\paragraph{Adjusting Modalities to Fit Available Tools and Capacity}
Participants also changed modalities when their availability or emotional capacity changed. A lower-effort or more available modality could keep an activity workable under different conditions. During bereavement, P16 moved from paper journals to a phone that was already nearby and less easily lost. When P02 felt too tired or weak to type, speech-to-text allowed an entry to be recorded by speaking. P10 used the Notes app when school left little time for handwriting, while P05 switched from a notebook to Apple Journal when away from home.

Physical modalities could be equally accessible when integrated into everyday settings. P14 kept a small booklet beside the computer and wrote three items while waiting for it to start. P11 similarly carried a small notebook that was ready whenever something came to mind. Convenience therefore depended on where a modality was available and how much effort it required at that moment.

Some participants combined modalities across stages of the same practice. P11 created entries on paper, then photographed and organized them by year in OneNote. This implementation preserved the expressive flexibility of paper while adding digital storage and search. Other participants moved between shorter digital entries and more detailed paper writing as their available time and energy changed.

Modality was therefore a flexible component of practice. Participants selected, shifted between, and combined modalities to align how they wanted to express an experience with what their circumstances allowed. These adjustments could preserve the purpose of an activity even as its observable form changed, showing why a change in modality did not by itself indicate disengagement.

\subsubsection{Rhythm: Fitting Practice into Daily Life}

\textbf{Participants developed rhythms that fit existing routines while allowing flexibility in when and how often they practiced}.

Rhythm describes how often participants practiced, when they practiced, and what prompted them to do so. Although all participants had long-term gratitude practices, their rhythms varied in frequency, regularity, and intensity. Some followed daily routines for years: P16 reported more than 1,780 consecutive days, while P15 had maintained a streak of approximately 530 days. Others practiced more flexibly. P10 might practice several times a week or return after two weeks to record something significant. Some \iquote{on and off}, for example described practicing\iquote{sometimes more intensely and sometimes less intensely for more than a decade.} 

Across this variation, participants organized practice around two recurring temporal patterns. Some embedded gratitude within daily routines, while others practiced when meaningful experiences made a situated psychological need salient. Many combined both: routines kept gratitude present in everyday life, while difficult or positive events prompted additional or more extended practice.

\paragraph{Using Daily Routines as Temporal Anchors}

Participants often attached practice to activities that already organized their day. Morning and evening anchors served different purposes: morning practices prepared attention for the coming day, as P15 explained:\iquote{How you start your day is how your day goes.} For many participants, evening practices supported reviewing and recording what had happened. P10 aimed to \iquote{sit down at the end of the day }and record \iquote{many lovely moments today}. P16 demonstrated the combination of both:
\begin{quote}
\iquote{I drink my first cup of coffee in the morning, and I am reading the gratitude prompt. \ldots In the evening, when I'm working through my wind-down, I'll reread the gratitude prompt, then I'll journal.}---P16
\end{quote}

For P16, the morning prompt provided a lens for noticing experiences during the day, and the evening created time to interpret and record them. Other participants used similar anchors. P14 read a daily text while drinking coffee to set a positive outlook for the day, and P12 placed gratitude alongside breakfast and French study.

These routines connected practice to existing activities and made it easier to sustain gratitude practice and position it at a time suited to its purpose. These anchors remained flexible: participants could still practice at other times or respond to events outside their routine.

\paragraph{Practicing Flexibly in Response to Meaningful Events}

Alongside scheduled routines, participants practiced when a situation produced a strong situated psychological need, which activated different patterns described in Section 4.1. Both difficult and positive situations could lead to an entry. P09 journaled when stress and bottled-up emotions became intense.P04 P07 wrote when something affected them emotionally.P02 recorded good news from work, friends, or family, while P17 wrote longer entries when a family or relationship experience moved him.

For some participants, rhythm changed as the life situation changed. P03 mentioned:

\begin{quote}
\iquote{I usually write when something emotionally significant happens. In 2024, I often wrote right after difficult moments, like a breakup, or feeling discouraged about work. It can be either in the morning, in the afternoon, [or] in the evening, whenever the situation happens.}---P03
\end{quote}

P03's rhythm shifted across years. In early 2024, her writing increased sharply during a period of hardship. By mid-to-late 2024, she wrote more steadily to work through ongoing tension. In 2025, positive milestones, such as moving out, beginning freelance work, and saving money also became reasons to write. By 2026, she described her practice as more stable and balanced.

Daily anchors and meaningful events could coexist within the same practice. Routines made practice available in everyday life, while event-triggered rhythms allowed participants to respond when particular experiences became salient. Rhythm therefore changed with the temporal purpose and situated relevance of practice; whether changes in regularity represented continued engagement or an interruption is examined in the next section.


\subsection{Sustaining Engagement Through Maintenance, Adaptation, Lapse, and Resumption}
\label{sec:continuity}
\begin{figure*}[t]
\centering
\includegraphics[width=\textwidth]{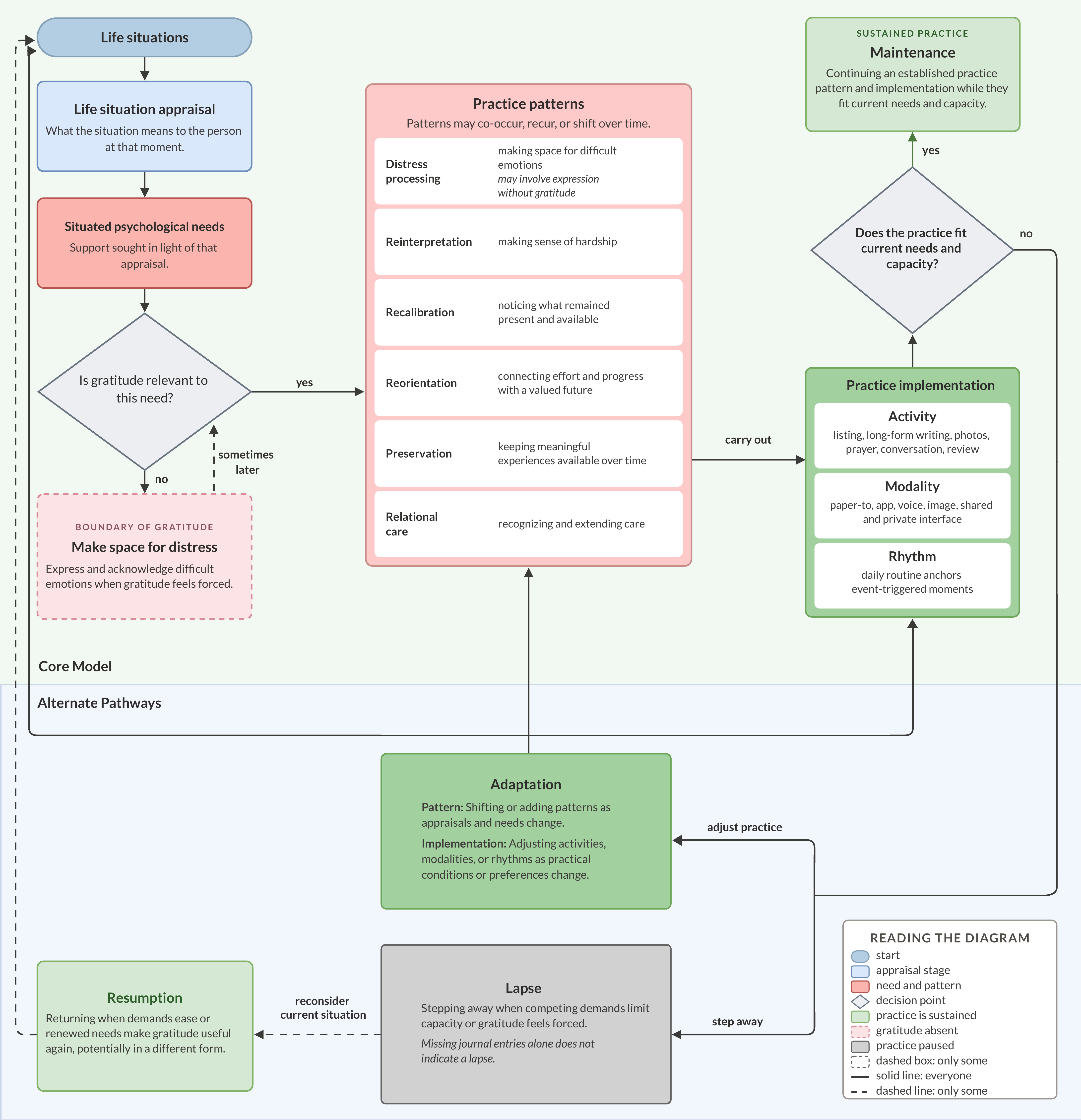}
\caption{This diagram shows the evolving adaptive model of long-term gratitude practice. Life situation appraisals shape situated psychological needs and whether gratitude feels relevant. Practice patterns are carried out through activities, modalities, and rhythms. When the implementation fits the need and capacity, it is repeated as maintenance. When the fit breaks down, people either adapt the pattern or the implementation and practice again, or lapse; practice remains available and is often resumed in a different form. }
\label{fig:study_detail}
\end{figure*}

We recruited participants with long-term gratitude practices to understand how they remained engaged with the gratitude practice over time. 
\textit{Maintenance} describes continued engagement through repeated use of an established practice pattern and implementation; \textit{adaptation} refers to practice that continued but whose patterns or implementation changed. \textit{Lapse} refers to stepping away from gratitude practice, and \textit{resumption} to returning after a lapse, potentially in a different form (Figure~\ref{fig:study_detail}). Missing journal entries alone did not indicate a lapse, as gratitude could continue through other activities or everyday awareness.

\begin{figure*}[t]
\centering
\includegraphics[width=\textwidth]{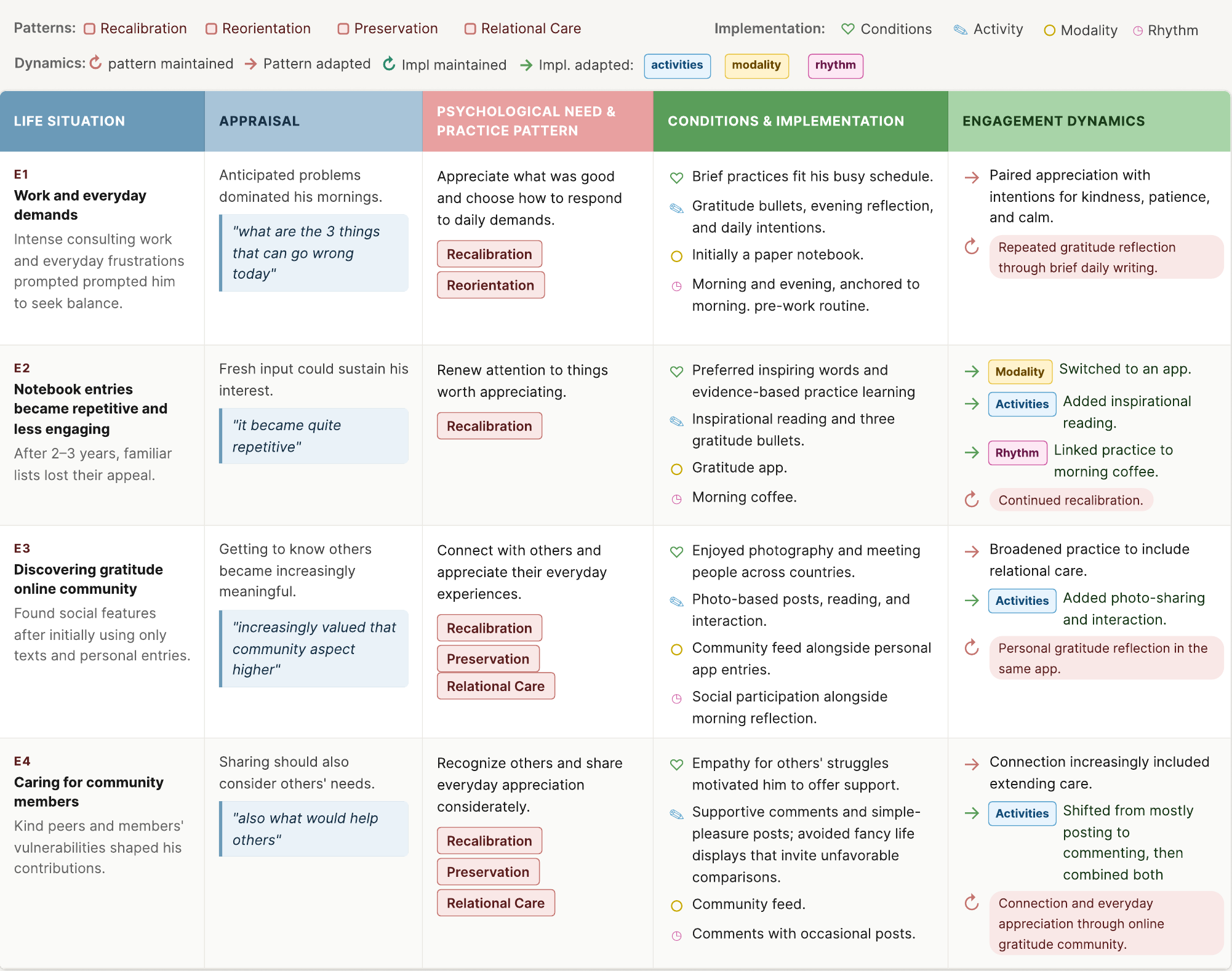}
\caption{P14's trajectory across four episodes (E1 -- E4). From personal reflection amid everyday demands to participation and care in an online community. The episodes show how repetitive notebook writing prompted a move to an app and how growing attention to others shaped his sharing and commenting. 
}
\label{fig:p14_example}
\end{figure*}

\begin{figure*}[t]
\centering
\includegraphics[width=\textwidth]{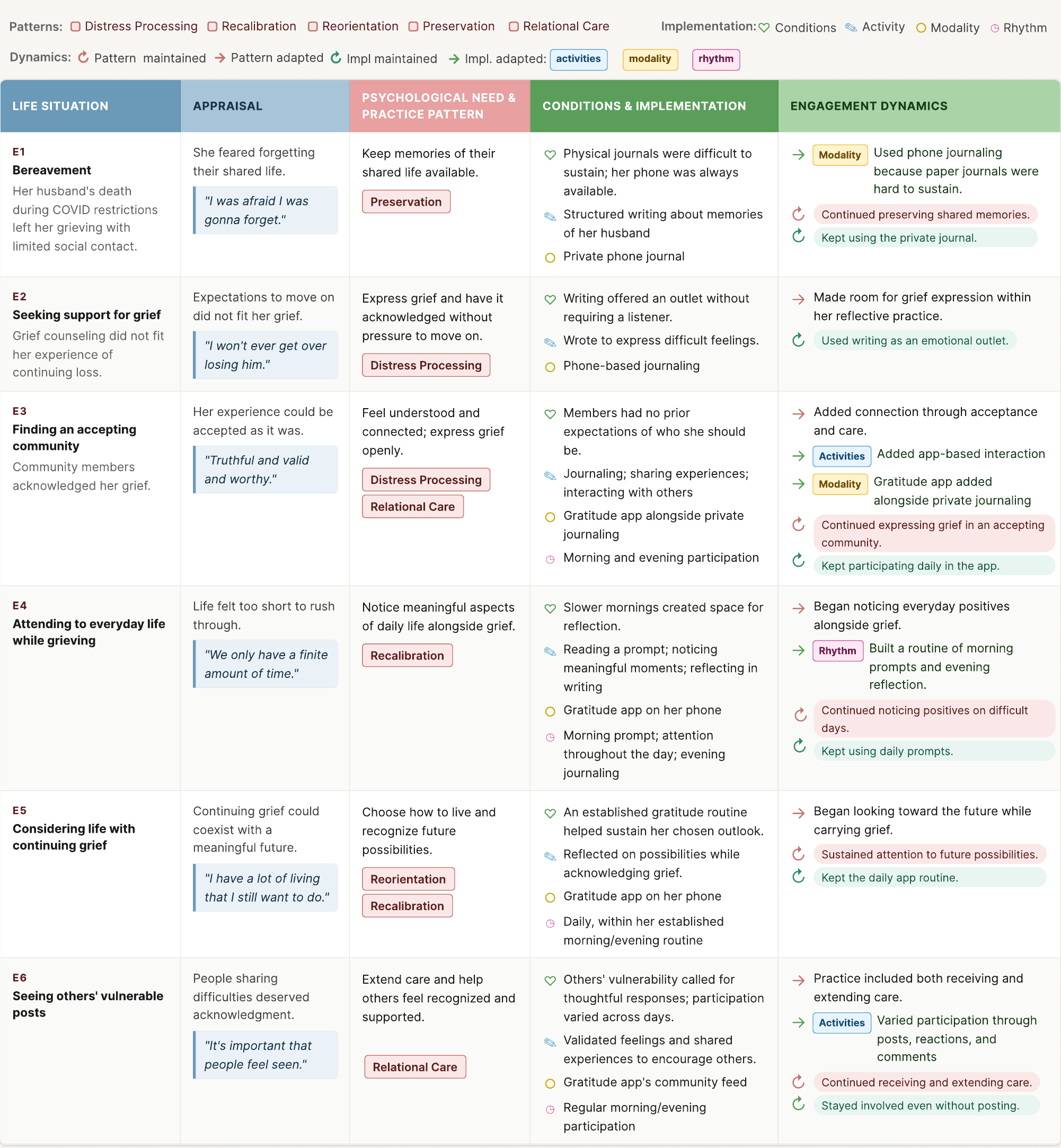}
\caption{P16's trajectory across six episodes (E1 -- E6). Following her husband's death. The episodes show how preserving memories, expressing grief, finding acceptance, appreciating everyday life, considering the future, and supporting others became relevant within ongoing bereavement.
}
\label{fig:p16_example}
\end{figure*}

We use P14's and P16's trajectories (Figures~\ref{fig:p14_example} and ~\ref{fig:p16_example}) to illustrate \textit{maintenance} and \textit{adaptation} within an individual's practice lifecycle. 
We describe this trajectory through \textit{episodes}. Each episode represents a snapshot of the flowchart (Figure~\ref{fig:study_overview}): the person has appraised their life situation and maintains practice pattern(s) that address their situated psychological needs, with an implementation that fits their practical conditions and preferences. A transition to a new episode begins when one of these construct changes or is reappraised, such as a new life situation contributing to adopting another pattern, or reappraising the situation and what is needed, or trying a different implementation.

We selected P14 and P16 to represent contrasting life situations. P14 practiced within a generally positive life with everyday work and family demands. P16 practiced through bereavement, illustrating major adversity also described by participants facing divorce, parental loss, or demanding caregiving. These situations and their appraisals shaped the psychological needs and patterns practice encompassed, while life practical conditions shaped its implementation.



\subsubsection{What Made Gratitude Worth Continuing or Resuming} Rather than relying solely on routine or immediate payoffs, participants sustained long-term engagement through two interlocking factors: \textbf{anchoring gratitude in broader personal commitments and recognizing tangible changes in how they navigated life}.

\paragraph{Participants Connected Gratitude to Broader Commitments}

\noindent Gratitude gained an ongoing place when participants connected it to valued identities, holistic well-being, and caring for others. For P16, gratitude had developed into a core value that guided her participation across different areas of life:

\begin{quote}
\iquote{When people ask me what my core values are, gratitude always shows up as one of those core values. Having a grateful heart is always one of the things that comes up for me, so I feel like that has become who I am as a person.}---P16
\end{quote}

For P16, gratitude had become part of how she understood the person she wanted to be. P14 similarly connected gratitude with being a healthy leader, parent, and friend. Participants also carried gratitude into how they supported others. P13 used gratitude prompt to help others feel comfortable and encouraged.

For other participants, gratitude became part of a broader commitment to holistic well-being. P07 combined mood reflection with tracking sleep, exercise, and food. P11, P12, P15, and P17 practiced gratitude alongside therapy, exercise, meditation, breathwork, or other reflective activities. Gratitude also supported addiction recovery. P01 used gratitude alongside meditation to regulate distress without returning to substance use.

These cases show that gratitude gained lasting value when it supported participants' efforts to care for their well-being, become the people they wanted to be, strengthen relationships, and pass care forward. These commitments did not prevent interruptions, but they gave participants enduring reasons to continue or return as their circumstances changed.

\paragraph{Perceived Changes Reinforced the Value of Continuing}

Participants also perceived changes in how they responded to difficulty, treated themselves, and related to others. Some recognized these differences through accumulated records, while others compared their current responses with how they remembered acting earlier in life. These retrospective accounts do not isolate gratitude as the cause of change. However, associating gratitude with personal growth gave the practice meaning beyond individual episodes and made it worth continuing. P03 described seeing this change across several years of journal entries:
\begin{quote}
\iquote{When I read all the entries from 2024, compared to 2025, and also now, I feel like I can see how my tone changed from feeling so lost, to feeling more confident, more independent. I feel reflection helps me respond to life instead of just reacting to it.}---P03
\end{quote}

For P03, comparing entries made a change in tone and self-understanding visible. Across participants' accounts, greater flexibility in responding to difficulty was a common thread connecting these perceived changes. P07 described moving away from persistent \iquote{what-if} thinking, while P15 became less focused on \iquote{worst-case scenarios} and more attentive to what remained present. P16 developed a \iquote{both-and mindset} that allowed grief and gratitude to coexist. P10 and P17 similarly described becoming more able to see the \iquote{glass half full} across different experiences.

This greater room to pause and reconsider a situation was also reflected in how participants treated themselves and others.  P12 became more willing to treat a setback as a chance to \iquote{start again} rather than be harsh on herself, while P11 became more intentional about expressing love, apologizing, helping family members, and thanking them for small things.

Participants did not describe the disappearance of difficulty or constant optimism. They perceived greater room to reconsider a situation and choose a more patient or caring response. Seeing these changes reinforced their sense that gratitude remained worth carrying forward.

\subsubsection{How Changing Capacity and Needs Shaped Lapse and Resumption}
\label{sec:lapse}
Participants \textbf{stepped away when competing demands limited their capacity or gratitude felt forced}, and \textbf{returned when demands eased or renewed needs made gratitude useful again}.
We use \textit{lapse} to describe periods when participants reported stepping away from gratitude practice. Missing entries alone did not establish a lapse. For example, P17 explained that practice did not have to remain \iquote{in a book,} since he still carried gratitude in his everyday awareness when travel disrupted his journaling. Switching tools or continuing inwardly could therefore sustain practice even when personal entries were absent. Participants described lapses associated with competing demands and times when gratitude felt emotionally inappropriate.

\paragraph{Competing Demands Limited Capacity to Practice}When participants had limited time, energy, attention, or access to their usual routine, their practices became difficult to carry out. P09 recalled pausing during college examinations:
\begin{quote}
“I had some kind of exam and stuff and midterms, I need to focus on studies. \ldots I did not journal.”---P09
\end{quote}
Looking back through her entries, she linked the lapse to competing demands without suggesting that gratitude had lost its personal value.

\paragraph{Emotional Unreadiness Made Gratitude Feel Forced}

Participants also stepped away when gratitude felt emotionally inauthentic. P07 still wanted to continue the practice, but on some days she ignored reminders because writing would feel forced:
\begin{quote}
“Some days I just ignore it, because I know that I wouldn't be able to write anything down, I would just be forcing myself.”---P07
\end{quote}
Her longer-term commitment coexisted with choosing not to practice on particular days. Practical and emotional difficulties could also overlap. P05 described a lapse during the pandemic and graduate-study pressures, when multiple demands reduced her motivation to reflect. Producing positive entries also felt like pretending:
\begin{quote}
“Unless I truly experience joy or happiness, it's hard for me to pretend that I'm writing something to make myself feel better.”---P05
\end{quote}
For P05, the demands on her attention and the difficulty of finding genuine appreciation jointly shaped her lapse.

\paragraph{Easing Demands and Renewed Needs Supported Resumption}
Participants resumed as demands eased or gratitude again seemed useful. P09 linked resumption to the end of her examinations. For P05, changing circumstances coincided with a renewed need for support. She felt off-balance after the pandemic and wanted to support herself again. She later restarted writing with her family, used reminders, and allowed flexibility in frequency and spoken expression. Her return combined renewed relevance with an adapted implementation.

P06 similarly described stopping writing when distressed, then returning when she wanted a \iquote{positive push.} Noticing small experiences, such as trying a new dish or speaking with a friend, gave her reasons to write again. These accounts suggest that distress could accompany both withdrawal and renewed engagement, depending on what participants sought and whether gratitude felt useful at the time.

Alongside changes in circumstances and psychological needs, the broader commitments and perceived changes described in Section 4.3.1 provided underlying reasons to return to gratitude practice. These accounts describe lapses within the histories of people who were practicing gratitude at interview; they offer limited insight into permanent discontinuation or why some people never resume.

\subsection{Participant-Envisioned Technology Support}
\label{sec:tech-support}
When asked how technology might better support their long-term gratitude practice, participants described three desired roles: responding to their current state, helping them interpret accumulated records to support self-understanding, and supporting appreciation within relationships. 

\subsubsection{Context-aware support: Responding to Current State Without Imposing Gratitude}
\label{sec:context}

Participants first wanted technology to consider their current emotional state before offering a gratitude-related support.This mattered most during distress, when they first wanted to express their feelings, understand what had happened, or calm down. P07 imagined an app that first asked about her mood before deciding whether gratitude fit the moment:
\begin{quote}
\iquote{I want an app that asks me about my mood, and not just ask 'What are you grateful for today?' \ldots There are days that the app should be able to evaluate and say, okay, today is not one of those good days that she can be grateful. \ldots It can say, 'What can you do to feel better right now?}'---P07
\end{quote}

When P07 was willing to reflect, she wanted the app to help her make sense of her distress,\iquote{ask reflective, thoughtful questions} based on what she had written, these questions could help her \iquote{find the root cause of why I'm feeling the way I'm feeling} because she often struggled to \iquote{piece [her] thoughts together.} P11 similarly imagined a conversational tool that could ask whether he wanted to talk and help him \iquote{get it out} before introducing gratitude. These accounts echoed the boundary identified in Section 4.1: participants wanted technology to check their readiness before presenting gratitude-related support.

The need for context also shaped reminders. Many participants relied on alarms and app notifications to maintain her routines, yet found the same reminders annoying on days when they felt low. P09 felt that guilt-based reminders would \iquote{ruin the fun in journaling.} P12 proposed a low-pressure message: \iquote{It's okay that you missed two days. You can always get started again.} Reminder preferences could therefore change across people and situations, and even within the same participant as their emotional state changed.

Together, these accounts describe context-sensitive support across two moments: checking whether gratitude fit the user's current emotional state and offering appropriate emotional or gratitude-related support, then responding to interruptions in ways that made returning feel manageable.

\subsubsection{Data-Supported Self-Understanding Over Time}
\label{sec:understand}
Participants wanted technology to help them learn from the journal records they had accumulated over time. By connecting entries across different periods, technology could help them recognize emotional patterns, see what had changed or remained unresolved, and use these insights to guide future goals and practices.P11 had already used AI in this way:

\begin{quote}
\iquote{I actually took all my journal entries from last year, dumped them into Microsoft Copilot, and got it to analyze it for me and spit out themes \ldots It'll tell you, Your mood seems like it was down at this point in time.}---P11
\end{quote}

P03 similarly imagined viewing emotional changes across different periods as “seeing my resilience on a timeline.” P02 focused more on comparing his mood and practice consistency across different periods so he could adjust his future practice.

Participants also wanted records to show which experiences had changed and which still needed attention.P08 imagined a report showing which difficulties had continued and which had been resolved.
P15 wanted to revisit major events and ask, “Have I worked through all this stuff? Do I still have things in there I need to figure out how to deal with?”  Records could therefore support reflection on both progress and unresolved experiences.

Some participants wanted these insights to guide future action.P12 imagined listing personal or mental-health goals, receiving possible ways to pursue them, and using prompts to review her daily progress. P17 similarly wanted general lessons to be \iquote{applied to your personal life} through journaling and reflection. Longitudinal support could therefore connect past experiences with current goals and future practice.

Together, participants envisioned longitudinal support that helped them recognize change, identify unresolved experiences, and connect past records to future action while preserving control over interpretation and resurfacing.


\subsubsection{Supporting Appreciation as Communication and Care}
\label{sec:care}

Participants further imagined technology helping them act on appreciation within relationships. This could involve prompting a message or conversation, connecting someone to a trusted person, or helping users respond more attentively to another person's needs. P11 described AI as a possible \iquote{gateway} to human support:

\begin{quote}
\iquote{AI could say, ‘Look, I can tell you're feeling down. Would you like me to call your friend Bob, or your friend Jim? Let's bring him in on the conversation, and then we can all talk about it.} ---P11
\end{quote}

In this account, technology did not replace relational support. It offered a route through which the user could choose to involve someone they trusted. P11 also imagined lightweight prompts such as, \iquote{Have you told her that you love her today?} P03 described a similar movement in her existing practice. Journaling helped her notice how friends and family had supported her, which sometimes led to “a message, a call, or just a small gesture.”

For P05, relational support began with listening and empathy. She described gratitude as a bonus \iquote{that could grow from effective conversation.} She imagined technology helping users understand what another person might be feeling and practice an appropriate response. P05 summarized this process as \iquote{listening, understanding, and perceiving, then providing feedback.} Her account extends relational support from prompting expressions of thanks to developing the communication skills that make appreciation genuine.

Together, these accounts position technology as a bridge from private experience to human care. Across the three forms of support, participants envisioned technology as a flexible resource that could respond to their current state, support self-understanding from accumulated records, and help them connect with others while preserving control over interpretation and sharing.

\section{Discussion}

Our findings characterize long-term gratitude as a situated and adaptive well-being practice.The \model{} shows how life situation appraisals shape psychological needs and practice patterns, while showing how practical conditions and preferences shape implementations. It captures engagement dynamics over time as needs and circumstances change. For HCI researchers and designers, it provides a framework for understanding why the same activity may feel useful in one situation and burdensome in another, and how adaptation, lapse, and resumption relate to changing life situations.

We derive three design implications by connecting participants' envisioned technology support (Section \ref{sec:tech-support}) with the findings in Sections (\ref{sec:relevance}--\ref{sec:continuity}): providing situated support, supporting self-understanding through past records, and facilitating relational care. Together, the model and implications guide the design and evaluation of gratitude technologies around how well support fits people's changing lives, alongside how often a system is used.

\subsection{Support Gratitude as a Situated and Evolving Practice}

Our findings show how hardship could be both a source and a boundary of gratitude. Participants sometimes found appreciation through difficulty, while other moments required acknowledgment and emotional expression before gratitude felt appropriate. This relationship aligns with second-wave positive psychology, research on posttraumatic growth and continuing distress~\cite{lomas2016second,dekel2012posttraumatic}, and gratitude-chatbot research showing how appreciation can arise within negative experiences~\cite{lee2024cultivating}. Emotional readiness therefore matters alongside the potential value of gratitude.

Participants' appraisals helped explain what support they sought. Similar situations could raise different needs, while different situations could raise similar needs, echoing multifinality and equifinality, respectively~\cite{cicchetti1996equifinality}. These relationships also changed within individuals over time. As circumstances and needs shifted, participants adjusted their practice patterns and implementations, or paused when gratitude felt burdensome or inappropriate. Broader commitments and perceived changes nevertheless gave some of them reasons to maintain and resume.

Personal informatics research similarly situates practice within everyday and social contexts~\cite{rooksby2014personal,ayobi2018flexible,murnane2018personal} and recognizes changing goals, lapses, and returns~\cite{epstein2015lived,ekhtiar2025changing}. Our findings connect these dynamics to gratitude's changing relevance and perceived value. Supporting this evolving practice involves understanding what fits the present moment and helping users reflect on how their needs and practices have changed. These complementary directions respond to participants' requests for situated support and longitudinal self-understanding (Sections \ref{sec:context}, \ref{sec:understand}).

\paragraph{Design Implication: Provide Situated Support}

To better provide situated support, systems could first invite users to share and revisit personal values and broader commitments, such as sustaining recovery or caring for family. This information could connect reflection with the commitments that made practice worth maintaining or resuming (Section\ref{sec:continuity}). Systems could then ask about users' recent circumstances and preferences, which could guide timing and interaction. For example, embedding practice in a morning routine, adjusting reminders during busy periods, or offering short lists, guided writing, and voice input to fit users' practical conditions and preferences (Section \ref{sec:implementation}).

Within an interaction, systems should address two questions: whether gratitude is appropriate now and, if so, what it should help the user do. Optional questions about what happened, what it means, and what would help could clarify current appraisals and needs. When users want to express distress, open-ended questions grounded in their account could help them clarify feelings and explore what makes an experience difficult (Section \ref{sec:context}). Those ready for gratitude could choose prompts for understanding hardship, noticing what remains available, appreciating effort, preserving memories, or recognizing care (Section \ref{sec:relevance}). These directions should remain open to combination and revision. When limited capacity or emotional unreadiness makes reflection burdensome, users should be able to skip questions or pause without explanation (Section\ref{sec:lapse}). Our findings highlight an important consideration for HCI: gratitude technologies should respect both the boundary of gratitude's relevance, supporting distress expression without requiring appreciation (Section \ref{sec:boundary}), and the boundary of engagement, allowing people to step away from gratitude practice (Section\ref{sec:lapse}).

\paragraph{Design Implication: Support Self-Understanding Through Past Records}

Participants wanted accumulated records to help them understand recurring concerns, recognize change, and revisit unresolved experiences (Section \ref{sec:understand}). User-requested summaries or timelines could connect related entries, helping users compare how they appraised similar situations and what they needed (Sections \ref{sec:relevance}, \ref{sec:implementation}).

Consider an illustrative scenario: a user asks the system to revisit entries about work difficulties. A timeline brings together earlier reflections focused on anticipated problems and later entries that also acknowledge effort and support, alongside the practices used and any comments on their usefulness. Reviewing these entries could help the user consider how their appraisals and needs changed, which practices felt helpful, and what remains unresolved. These insights could inform how they approach similar situations and adapt their future practice.

Recognizing changes they value could help users reflect on what makes practice worth continuing or resuming (Section \ref{sec:continuity}). Summaries should link interpretations to original entries so users can assess their relevance, add context, or reject them. Users should also control which periods and experiences are included or resurfaced. Insights they endorse could inform future situated support, subject to their current needs and readiness.

\subsection{Support Relational Care Across Gratitude Practices}

Our findings connect relational gratitude with participants' situated needs for reassurance, belonging, and opportunities to support others. Private reflection also helped participants recognize care they had received, sometimes prompting them to thank, encourage, or help others. These accounts show how receiving care, recognizing it, and extending it could become connected within everyday practice.

This finding builds on research describing gratitude cycles in online communities \cite{makri2020can} and the role of gratitude in recognizing responsive relationship partners and strengthening social bonds \cite{algoe2012find}. Our findings highlight how these relationships mattered during experiences such as grief, shame, and isolation. In these circumstances, acknowledgment of a person's difficult experience could itself become a source of gratitude. Participants' accounts therefore suggest supporting care across moments of seeking support, reflecting privately, and responding to others.

\paragraph{Design Implication: Help Users Seek and Express Care.}

Participants envisioned technology helping them reach trusted people, express appreciation, and respond thoughtfully to others (Section~\ref{sec:care}). Systems could help users contact someone they trust when support is wanted, or revisit care recorded in their journals and consider expressing appreciation through a message, call, or gesture. These features could connect private reflection with interpersonal care.

Participants also emphasized listening and perspective-taking. Questions or suggested responses could help users consider another person's experience and prepare a response that acknowledges their difficulty or offers support. AI feedback can potentially help peer supporters respond more empathically in text-based mental health support~\cite{sharma2023human,kim2026llumi}, suggesting a direction to explore in gratitude communities.

Users should retain authority over what they share and how they respond. AI suggestions should remain optional and editable, and receiving care should carry no expectation of gratitude in return. This allows appreciation to develop through supportive interaction while leaving users free to express distress, accept support, or offer care on their own terms.

\subsection{Limitations and Future Work}

Our study has two main limitations. First, participants recalled practices spanning one to fifteen years, making their accounts subject to recall bias. Revisiting gratitude records helped ground interviews in concrete examples, but current perspectives and selective recall may still have shaped how participants described past experiences and changes. Future longitudinal research could combine repeated interviews with in-situ accounts to examine how appraisals, needs, and practices change as life situations unfold. 

Second, we deliberately recruited people with long-term gratitude practices to understand how they navigated changing life situations and to inform technology that could support sustained practice more broadly. Participation was also self-selected: we studied people who already engaged in gratitude journaling and chose to participate. This sampling strategy provided rich accounts of maintenance, adaptation, lapse, and resumption, but may not capture the experiences of people who do not journal or who permanently discontinued the practice. Our account of change therefore reflects the histories of people who continued or returned to practice and cannot explain all trajectories of disengagement. Future research could examine why people do not practice gratitude, what pain points they encounter, and how technologies could be designed to support these non-users.

\section{Conclusion}

Gratitude technologies can struggle to accommodate people's changing life situations: prompts may feel intrusive or create pressure when gratitude does not fit their current needs or emotional readiness. Our \model{} explains how participants' appraisals shaped whether gratitude felt appropriate, what they sought from it, and how they implemented practice. Long-term practice involved maintenance, adaptation, lapse, and resumption as needs and capacity changed, while perceived value and broader life commitments provided reasons to continue or return (RQ1). Participants envisioned technology that responds to current needs, supports self-understanding through past records, and facilitates appreciation and relational care (RQ2). These insights inform gratitude technologies that support meaningful practice across changing circumstances, including when people need to adapt their practice, pause, or seek other forms of support.

\begin{acks}
This work was supported in part by the National Institute on Aging of the National Institutes of Health under Award Number P30AG073105 and the Jump ARCHES endowment through the Health Care Engineering Systems Center at the University of Illinois and the OSF Foundation.
\end{acks}

\bibliographystyle{ACM-Reference-Format}
\bibliography{main}


\appendix
\clearpage
\appendix
\begingroup
\raggedbottom

\section{Appendix}

This appendix presents examples of participants' gratitude records, illustrating what these artifacts looked like and grounding our findings in the artifact-elicitation interviews. Table~\ref{tab:artifact-index} links each example to the relevant finding.

\subsection{Participants' Journal Examples}
\label{app:entry}

\begin{table}[!htbp]
\centering
\small
\caption{Guide to the appendix materials and their connections to the findings.}
\label{tab:artifact-index}
\begin{tabular}{@{}p{0.15\linewidth}p{0.34\linewidth}p{0.42\linewidth}@{}}
\toprule
\textbf{Figure} & \textbf{Gratitude Entries} & \textbf{Connection to findings} \\
\midrule
Fig.~\ref{fig:P03} & P03: journal reflections & Reinterpreting difficulty \\
Fig.~\ref{fig:P10} & P10: entry from a family trip & Preserving meaningful experiences \\
Fig.~\ref{fig:P11} & P11: exercise calendar & Reorientation; visible records of effort \\
Fig.~\ref{fig:P02} & P02: displayed gratitude stickers & Relational assurance; visible records \\
\bottomrule
\end{tabular}
\end{table}

\begin{figure}[!htbp]
    \centering
    \includegraphics[width=0.48\linewidth]{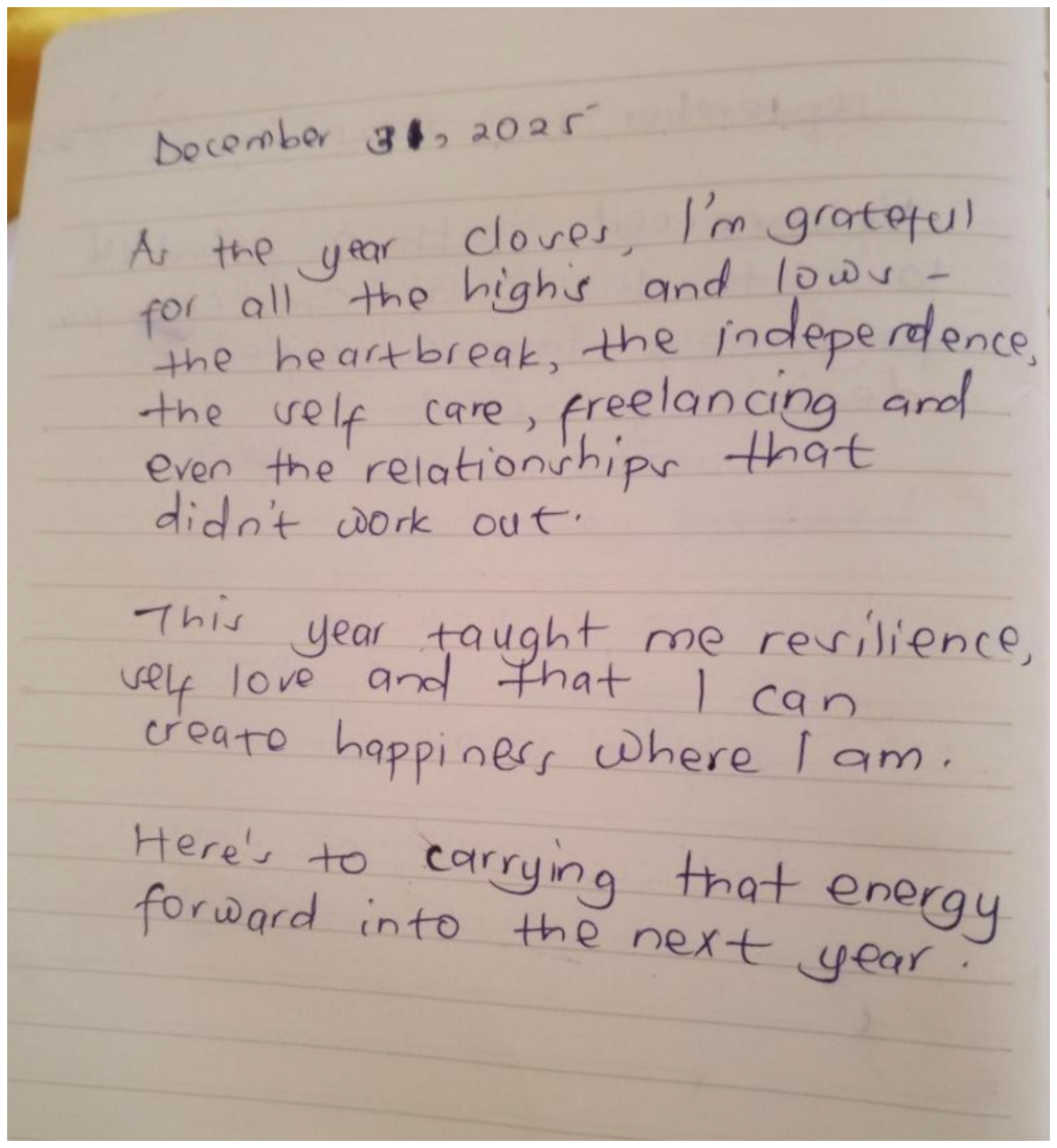}
    \caption{P03's journal reflections during a breakup and prolonged unemployment. Rather than framing the breakup or job loss as good things, P3 asked what they taught her: ``I realized I was actually being resilient and just trying consistently... I don't turn bad into good, I just look for growth inside.'' The entry accompanies the finding on reinterpreting difficulty.}
    \Description{A handwritten journal entry reflecting on a breakup and unemployment, considering lessons learned, resilience, and personal growth.}
    \label{fig:P03}
\end{figure}

\begin{figure}[!htbp]
    \centering
    \includegraphics[width=0.48\linewidth,height=0.58\textheight,keepaspectratio]{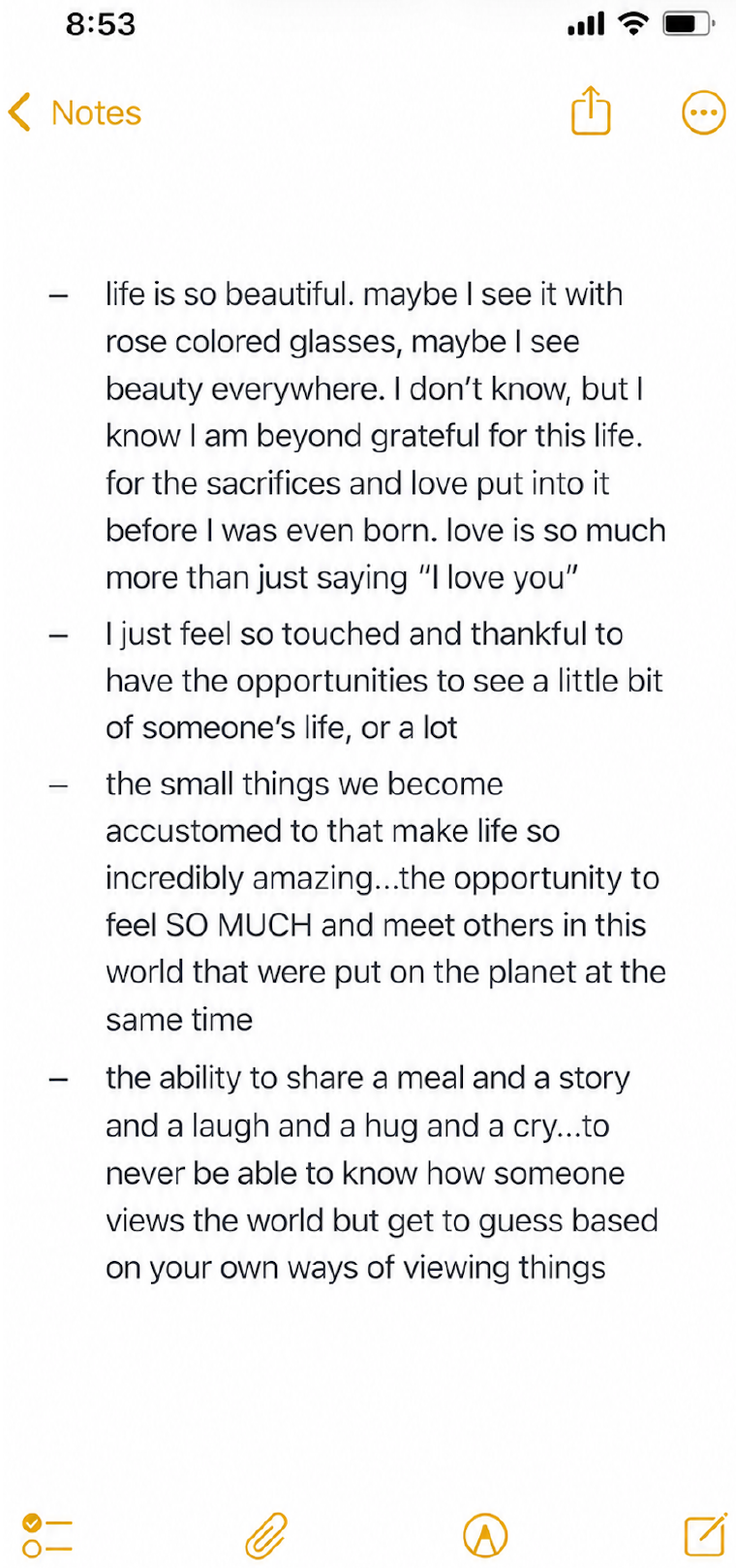}
    \caption{P10's gratitude entry from a freshman-year trip with her parents. P10 connected everyday moments to her family history, including her parents' immigration and her grandfather's survival during the Holocaust, describing it as a reminder she ``could have not been here at all'': ``The small energy moments are the things in our day-to-day... the ability to share a meal and a story and a laugh and a hug and a cry.'' The entry accompanies the finding on preserving meaningful experiences.}
    \Description{A gratitude entry from a family trip, reflecting on life, inherited love and sacrifice, and shared everyday experiences.}
    \label{fig:P10}
\end{figure}

\begin{figure}[!htbp]
    \centering
    \includegraphics[width=0.7\linewidth,height=0.32\textheight,keepaspectratio]{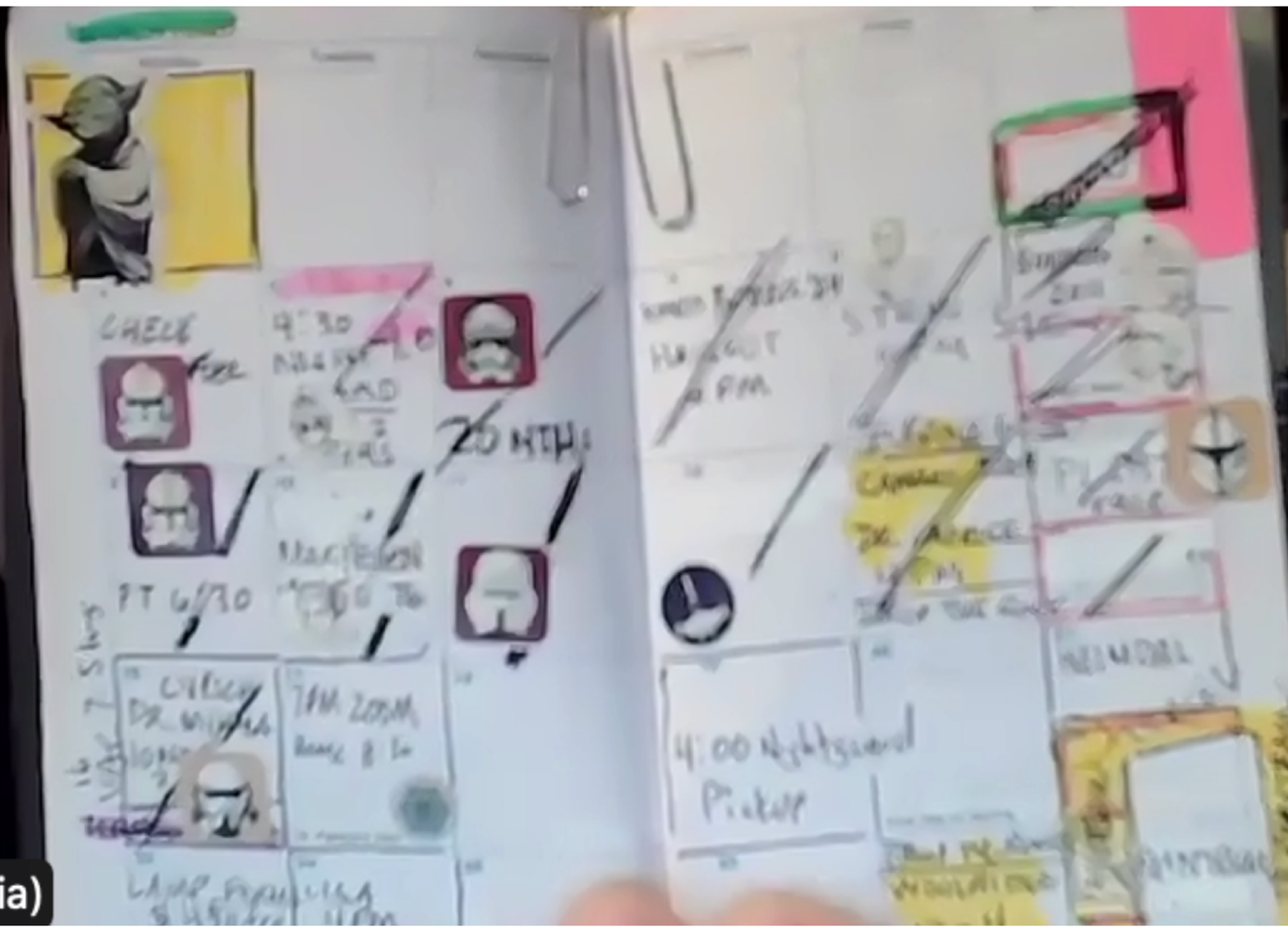}
    \caption{P11's calendar uses square stickers to mark days of exercise, making repeated effort visible over time. This visible record supported reorientation, letting him recognize small daily wins as progress toward a valued future.}
    \Description{A calendar with square stickers marking days when P11 exercised.}
    \label{fig:P11}
\end{figure}

\begin{figure}[!htbp]
    \centering
    \includegraphics[width=0.48\linewidth,height=0.58\textheight,keepaspectratio]{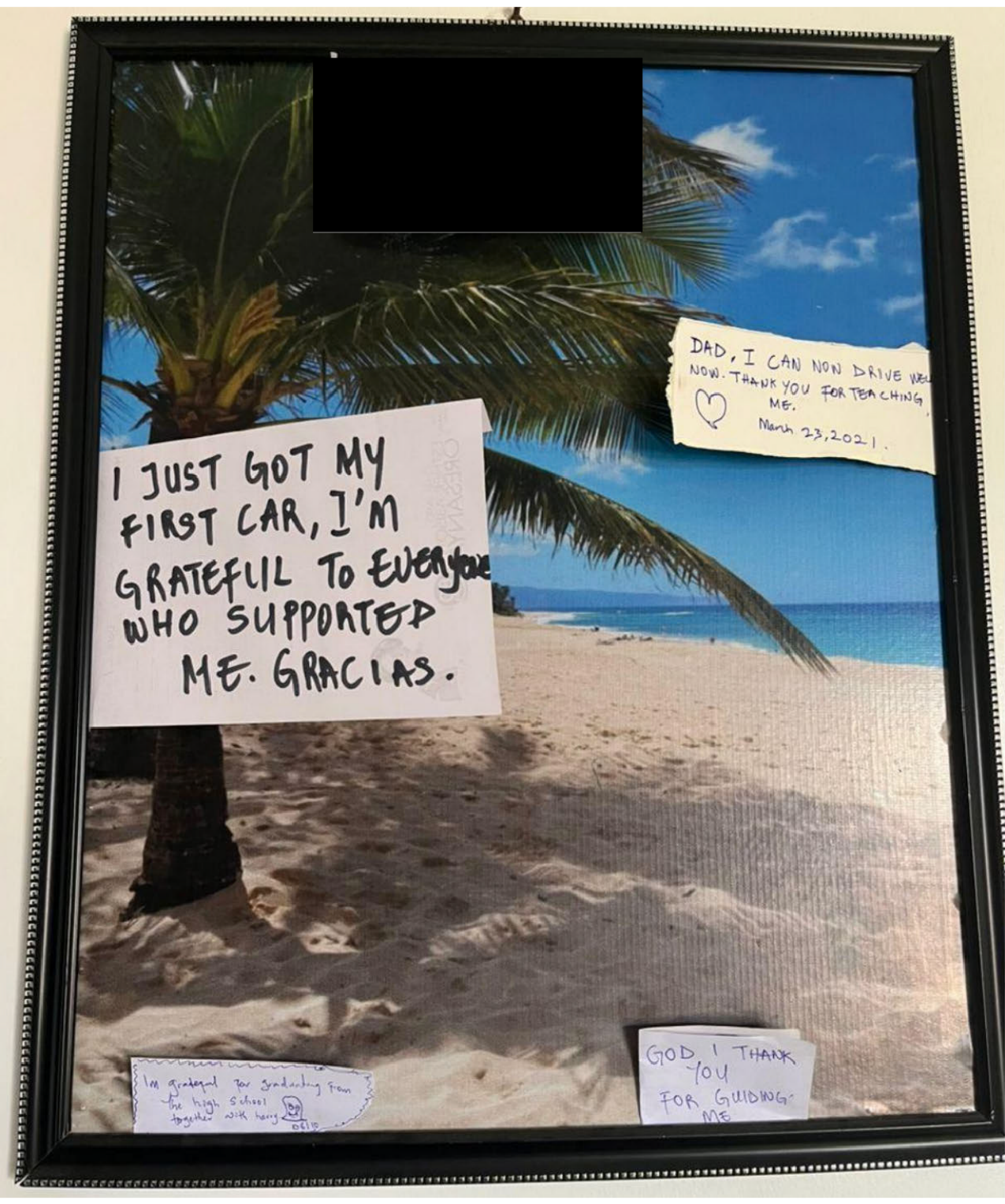}
    \caption{P02's displayed handwritten gratitude stickers. P2 kept these in view as a morning reminder of others' care: ``Anytime I woke up in the morning and saw those stickers, [they] made me feel happy, and made me feel like people out there were supporting me and really loved me.'' The material connects relational assurance with the everyday visibility of physical records.}
    \Description{Handwritten gratitude stickers displayed together, visible as part of a morning routine.}
    \label{fig:P02}
\end{figure}

\end{document}